\documentclass[sigconf]{acmart}
\usepackage{threeparttable}
\usepackage{color, xspace}
\usepackage{booktabs} % For formal tables
\usepackage{amsmath}
\usepackage{amsfonts}
\usepackage{comment}
\usepackage{footmisc}
\usepackage{mathrsfs}
\usepackage{graphicx}
\usepackage{subfigure}
\usepackage{multirow}
\usepackage{algorithm}
\usepackage{algorithmic}
\usepackage{multicol}  
\usepackage{enumitem}
\usepackage{array}
\usepackage{float}
\usepackage{hyperref}

\usepackage{soul, color, xcolor,xspace}
\newcommand{\name}{MTORL\xspace}

\copyrightyear{2025}
\acmYear{2025}
\setcopyright{acmlicensed}\acmConference[KDD '25]{Proceedings of the 31st ACM SIGKDD Conference on Knowledge Discovery and Data Mining V.2}{August 3--7, 2025}{Toronto, ON, Canada}
\acmBooktitle{Proceedings of the 31st ACM SIGKDD Conference on Knowledge Discovery and Data Mining V.2 (KDD '25), August 3--7, 2025, Toronto, ON, Canada}
\acmDOI{10.1145/3711896.3737250}
\acmISBN{979-8-4007-1454-2/2025/08}

\begin{document}

\title{Multi-task Offline Reinforcement Learning for Online  Advertising in Recommender Systems}
% \title{Offline reinforcement learning via multi-task GRU-Attention model for advertising}

% \author{Langming Liu\textsuperscript{1}, Wanyu Wang\textsuperscript{1}, Chi Zhang, Bo Li, Hongzhi Yin, Xuetao Wei, Wenbo Su, Bo Zheng, Xiangyu Zhao}
% \affiliation{%
%   \institution{\textsuperscript{1}Taobao \& Tmall Group of Alibaba, \textsuperscript{2}City University of Hong Kong, \textsuperscript{3}Harbin Engineering University, \textsuperscript{4}University of Queensland, \textsuperscript{5}Southern University of Science and Technology}
%   % \city{Hangzhou}
%   \country{}
% }

%1
\author{Langming Liu}
\authornote{Both authors contributed equally to this research.}
% \email{Langmiliu2-c@my.cityu.edu.hk}
\affiliation{%
  \institution{Taobao \& Tmall Group of Alibaba City University of Hong Kong}
  % \streetaddress{P.O. Box 1212}
  \city{Hangzhou}
  \country{China}
}

%2
\author{Wanyu Wang}
\authornotemark[1]
\affiliation{%
  \institution{Southern University of Science and Technology, City University of Hong Kong}
  \city{Hong Kong}
  \country{China}
}

%3
\author{Chi Zhang}
\affiliation{%
  \institution{City University of Hong Kong}
  \city{Hong Kong}
  \country{China}
}

%4
\author{Bo Li}
\affiliation{%
  \institution{City University of Hong Kong}
  \city{Hong Kong}
  \country{China}
}

%5
\author{Hongzhi Yin}
\affiliation{%
  \institution{University of Queensland}
  \city{Brisbane}
  \country{Australia}
}

%6
\author{Xuetao Wei}
\authornotemark[2]
\affiliation{%
  \institution{Southern University of Science and Technology}
  \city{Shenzhen}
  \country{China}
}

%7
\author{Wenbo Su}
\affiliation{%
  \institution{Taobao \& Tmall Group of Alibaba}
  \city{Beijing}
  \country{China}
}

%8
\author{Bo Zheng}
\affiliation{%
  \institution{Taobao \& Tmall Group of Alibaba}
  \city{Beijing}
  \country{China}
}

%9
\author{Xiangyu Zhao}
\authornote{Corresponding Authors}
\affiliation{%
  \institution{City University of Hong Kong}
  \country{Hong Kong}
  \country{China}
}

\renewcommand{\shortauthors}{Langming Liu, et al.}
\begin{abstract}

Online advertising in recommendation platforms has gained significant attention, with a predominant focus on channel recommendation and budget allocation strategies. 
However, current offline reinforcement learning (RL) methods face substantial challenges when applied to sparse advertising scenarios, primarily due to severe overestimation, distributional shifts, and overlooking budget constraints.
To address these issues, we propose MTORL, a novel multi-task offline RL model that targets two key objectives. 
First, we establish a Markov Decision Process (MDP) framework specific to the nuances of advertising. Then, we develop a causal state encoder to capture dynamic user interests and temporal dependencies, facilitating offline RL through conditional sequence modeling. 
Causal attention mechanisms are introduced to enhance user sequence representations by identifying correlations among causal states. We employ multi-task learning to decode actions and rewards, simultaneously addressing channel recommendation and budget allocation. Notably, our framework includes an automated system for integrating these tasks into online advertising. 
Extensive experiments on offline and online environments demonstrate MTORL's superiority over state-of-the-art methods. 
The code is available online at~\url{https://github.com/Applied-Machine-Learning-Lab/MTORL}.

\end{abstract}

\keywords{Advertising, Offline Reinforcement Learning, Multi-task Learning}

% 	<ccs2012>
% 	<concept>
% 	<concept_id>10002951.10003317.10003347.10003350</concept_id>
% 	<concept_desc>Information systems~Recommender systems</concept_desc>
% 	<concept_significance>500</concept_significance>
% 	</concept>
% 	</ccs2012>
% \ccsdesc[500]{Information systems~Recommender systems}

% <concept_id>10002951.10003260.10003272</concept_id>
% <concept_desc>Information systems~advertising</concept_desc>
% \ccsdesc[500]{Information systems~advertising}

\begin{CCSXML}
	<ccs2012>
	<concept>
	<concept_id>10002951.10003317.10003347.10003350</concept_id>
	<concept_desc>Information systems~Recommender systems</concept_desc>
	<concept_significance>500</concept_significance>
	</concept>
	</ccs2012>
\end{CCSXML}

\ccsdesc[500]{Information systems~Recommender systems}

\maketitle

\section{Introduction}
\label{sec:intro}
In recent years, advertising has emerged as a prominent approach for promoting products and services across various domains, encompassing e-commerce, short-video platforms, social media, and insurance~\cite{zhou2023direct,ku2023staging,pan2019predicting,avadhanula2021stochastic,gao2025generative}. Unlike traditional advertising, which confines promotion to fixed channels, online advertising systems offer dynamic and personalized ad placement policies~\cite{fan2007selling,danaher2010optimal} for online advertisers.
At the core of online advertising lies channel recommendation and budget allocation. \textbf{Channel recommendation} aims to autonomously identify suitable advertising channels (e.g., search ads, display ads, short videos, social media) for each exposure based on user preferences~\cite{zantedeschi2017measuring,zhang2009retailers}. Conversely, judicious \textbf{budget allocation} at the channel and other levels becomes imperative to maximize revenue~\cite{alon2012optimizing}.
While prior works have strived to elaborate corresponding advertising strategies~\cite{zhang2009retailers,de2016effectiveness,balseiro2020dual,naor2018near}, the proliferation of users and user features presents a formidable challenge for traditional approaches in accurately capturing user preferences, jeopardizing advertising performance.

% In recent years, advertising has emerged as a prominent approach for promoting products and services across various domains, encompassing e-commerce, short-video platforms, social media, and insurance~\cite{morris2009understanding,zhou2023direct,ku2023staging,pan2019predicting,avadhanula2021stochastic}. Unlike traditional advertising, which confines promotion to fixed channels, online advertising systems offer dynamic and personalized ad placement policies~\cite{fan2007selling,danaher2010optimal} for online advertisers.
% At the core of online advertising lies channel recommendation and budget allocation. \textbf{Channel recommendation} aims to autonomously identify suitable advertising channels (e.g., search ads, display ads, short videos, social media) for each exposure based on user preferences~\cite{rangaswamy2005opportunities,zantedeschi2017measuring,zhang2009retailers}. Conversely, judicious \textbf{budget allocation} at the channel and other levels becomes imperative to maximize revenue~\cite{alon2012optimizing,fischer2011practice}.
% While prior works have strived to elaborate corresponding advertising strategies~\cite{zhang2009retailers,dinner2014driving,de2016effectiveness,balseiro2020dual,naor2018near,buchbinder2007online}, the proliferation of users and user features presents a formidable challenge for traditional approaches in accurately capturing user preferences, jeopardizing advertising performance.

% \textbf{Concern: The authors ignore non-RL literature.}
Deep learning (DL) has revolutionized the advertising landscape~\cite{zhao2018deep1,cheng2016wide,zhou2018deep,zhao2020whole, zhao2022adaptive, zhao2021usersim, zhao2023user, liu2024modeling, lin2023autodenoise, wang2023plate, li2023hamur}, especially \textbf{channel recommendation}, leveraging deep networks to extract features from intra- and inter-channel across various ads and provide recommendation results through suitable channels~\cite{hao2021re,xu2023multi,liu2022neural}. However, most DL-based methods lack dynamic decision-making capabilities and ignore long-term rewards~\cite{kaelbling1996reinforcement,afsar2022reinforcement}. 
Deep reinforcement learning (DRL)~\cite{afsar2022reinforcement,wiering2012reinforcement} provides the solution to address the issues above, as its nature is to solve the MDP problem, making dynamic decisions to maximize long-term revenue~\cite{zhao2018deep, zhao2017deep, zhao2018recommendations, zhao2021dear, zhao2019deep, liu2023multi, zhang2020deep, zhao2020jointly}.
However, directly deploying untrained DRL into the online-serving advertising system is destructive for revenue and user experience~\cite{zhao2018deep}. Therefore, pre-training of DRL on the offline dataset is necessary for painless deployment, where the paradigm of offline RL~\cite{kostrikov2021offline,kumar2020conservative,kumar2019stabilizing} perfectly adapts to this environment. Unfortunately, the issues of overestimation and distributional shifts frequently emerge in offline environments. 
Inspired by existing advancements in the offline RL community~\cite{kumar2019stabilizing,agarwal2020optimistic}, some works strive to address such issues in advertising scenarios~\cite{liu2025session,korenkevych2024offline,kiyohara2021accelerating} by constraining the policy to adhere to the data distribution. 
However, this approach proves unsuitable due to the heightened sparsity of rewards in advertising datasets compared to general RL scenarios~\cite{chen2021survey,xiao2019model,shi2019virtual}.
Another line of offline RL methods~\cite{hussein2017imitation,ho2016generative} treats states as features and actions as labels, lifting the training efficiency by applying a sequential modeling paradigm, e.g., recent prevailing Transformer-based models~\cite{chen2021decision,wang2023causal}, including decision Transformer (DT). 
Despite the success of this approach, a new challenge arises: Transformer-based models do not adhere to the Markov property and exhibit suboptimal performance in capturing the temporal dependencies inherent in advertising (i.e., short-term data)~\cite{zeng2023transformers}.

In addition to the above challenges, existing works fall short in addressing critical considerations of \textbf{budget allocation} in advertising. One prominent approach is multi-touch attribution (MTA)~\cite{ji2017additional,ren2018learning,kumar2020camta}, a data-driven DL-based method that attributes contributions to individual touchpoints and channels. Subsequently, the attribution scores serve as the basis for further budget allocation. However, it is noteworthy that MTA methods lack dynamic decision-making capabilities since they merely consider static allocation instead of modeling user dynamic preferences~\cite{ji2017additional,kumar2020camta}.
Subsequently, constrained RL~\cite{ge2021towards,liao2022cross,cai2023marketing,zhang2021bcorle,cai2023two} is introduced in advertising to address the limits of MTA methods in dynamic decision-making, wherein budget limitation is incorporated as constraints within the constrained MDP (CMDP) framework. Specifically, the Lagrangian multiplier is applied to relax the problem into an unconstrained or soft-constrained one. 
Nevertheless, existing methods neglect dynamically selecting target users for advertising to satisfy the budget constraint, thereby overlooking an essential feature of advertising: a few users with high conversion tendencies can yield more returns than many users with low conversion tendencies~\cite{hao2020dynamic}. Moreover, users' conversion tendencies vary over time~\cite{lee2012estimating}.

In response to the aforementioned challenges, we present a novel multi-task offline RL model, \name, specifically tailored for addressing advertising concerns encompassing channel recommendation and budget allocation. 
We first formulate the advertising problem within the framework of offline RL. Subsequently, we employ a sequence modeling strategy to tackle the offline RL problem, where we significantly designate actions and rewards as labels to rich supervised signals, addressing the influence of overestimation and distributional shifts. In this context, we propose a causal state encoder~\cite{bai2018empirical} to capture temporal dependencies. 
Furthermore, we introduce the causal attention module~\cite{vaswani2017attention,radford2018improving} to enhance prior information collection in the sequence.
The enhanced sequence representation is directed into two branches of multi-task learning: one devoted to the action decoder, directly influencing the channel recommendation policy, and the other to the reward decoder. We also propose direct preference optimization (DPO) loss to mitigate issues caused by highly sparse rewards. 
The reward predictions play a pivotal role in channel- and user-level budget allocation. The automated integration of channel recommendation and budget allocation modules guides the advertising procedure. 

% In response to the aforementioned challenges, we present a novel multi-task offline RL model, \name, specifically tailored for addressing advertising concerns encompassing channel recommendation and budget allocation. 
% Our approach is initiated by formulating the advertising problem within the framework of offline RL.  
% Subsequently, we employ a sequence modeling strategy to tackle the offline RL problem, where, significantly, we designate actions and rewards as labels representing channel recommendation and budget allocation, respectively. In this context, we propose a causal state encoder~\cite{bai2018empirical} designed to capture temporal dependencies. 
% Furthermore, we introduce the causal attention module~\cite{vaswani2017attention,radford2018improving}, featuring a self-attention mechanism equipped with a causal mask. This module serves to collect prior information to enhance user sequence representation learning.
% The learned sequence representation is directed into two branches of multi-task learning: one devoted to the action decoder, directly influencing the channel recommendation, and the other to the reward decoder. 
% Subsequently, the reward predictions derived from our model play a pivotal role in budget allocation, especially in selecting target users. We establish a comprehensive and automated channel- and user-level budget allocation procedure. 

The major contributions to our work are summarized as follows:
\begin{itemize}[leftmargin=*]
\item We formulate the channel recommendation and budget allocation in advertising into a well-structured offline RL problem, where each fundamental concept within advertising is rigorously defined within the framework of CMDP. 
\item We propose an innovative multi-task offline RL model (\name) for channel recommendation and budget allocation. In addition, we develop an automated advertising procedure that integrates channel recommendation and budget allocation modules. This practical framework offers an efficient solution for advertising.
\item Our work encompasses extensive experiments on two public benchmark datasets. % and practical commercial datasets. 
The results unequivocally demonstrate the superior performance of \name compared to other state-of-the-art baselines. 
We also conduct online experiments to validate its effectiveness in the online environment.
\end{itemize}

% \begin{itemize}[leftmargin=*]
% \item We formulate the channel recommendation and budget allocation in advertising into a well-structured offline RL problem, where each fundamental concept within advertising is rigorously defined within the framework of CMDP. 
% \item We propose an innovative multi-task offline RL model (\name) for channel recommendation and budget allocation. \name incorporates a causal state encoder capable of capturing temporal dependencies and the user's recent preferences, while the causal attention captures the correlation between causal states. 
% \item We develop a comprehensive and automated advertising procedure that integrates the channel recommendation and budget allocation modules, where the budget allocation operates at both channel and user levels. This practical framework offers a convenient and efficient solution for online advertising.
% \item Our work encompasses extensive experiments conducted on public benchmark datasets, namely Kuairand and Criteo. The results unequivocally demonstrate our proposed \name model's superior performance compared to other state-of-the-art baselines.
% \end{itemize}
\begin{figure*}
    \centering
    \includegraphics[width=1.0\linewidth]{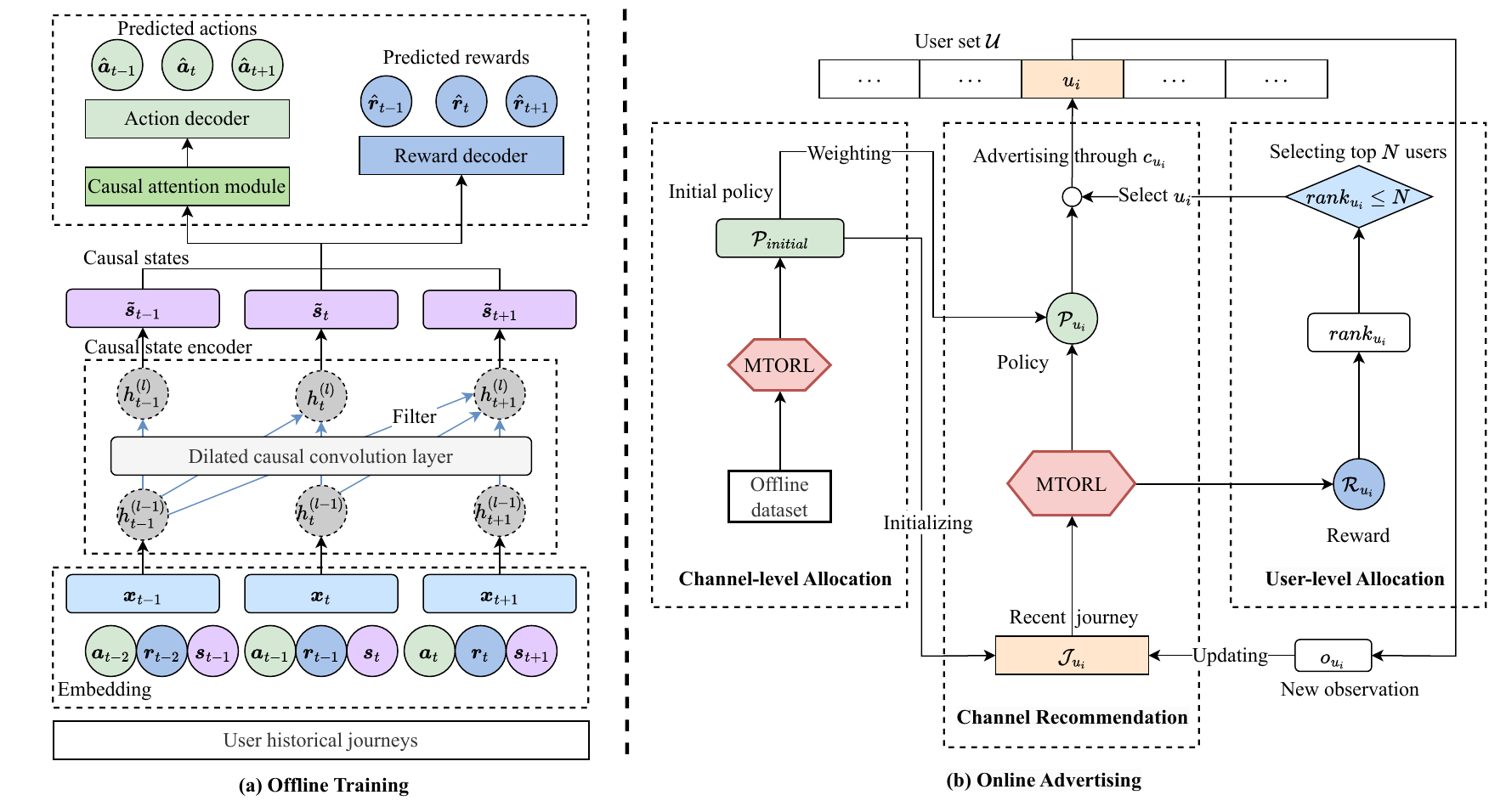}
    % \caption{Architecture}
    % \vspace{-7mm}
    \caption{The overview of MTORL. Part (a) shows the network architecture, which inputs the historical journeys and outputs predicted actions and rewards. Part (b) demonstrates the workflow of using MTORL in online advertising, where the channel recommendation module generates ad exposure policy. The channel- and user-level allocation modules control the ad exposures by reweighting policy and selecting target users, respectively. }
    \label{fig:Architecture}
    % \vspace{-4mm}
\end{figure*}
% The overview of MTORL. Part (a) shows the architecture of the network of MTORL, which inputs the historical journeys and outputs predicted actions and rewards. MTORL is trained on an offline dataset using the optimization method in Section~\ref{sec:opt}. Part (b) demonstrates the workflow of using MTORL in online advertising, where the channel recommendation module generates ad exposure policy for users. The channel- and user-level allocation modules control the ad exposures by reweighting policy and selecting target users, respectively. 

\section{Preliminary}
\label{sec:preliminary}
This section introduces the necessary preliminaries and formally presents the problem of multi-task offline RL for advertising.
% In this section, we introduce preliminaries and critical definitions, and then formally present the studied problem of multi-task offline reinforcement learning for advertising in this work.

\subsection{Multi-channel Advertising}
\label{sec:ad}
Given a set of users $\mathcal{U}=\{u_1,u_2,\cdots,u_{\vert\mathcal{U}\vert}\}$ and a set of channels $\mathcal{C}=\{{c}_1,{c}_2,\cdots,{c}_{\vert\mathcal{C}\vert}\}$, denote $u_i$'s static feature (e.g., user profiles) as $\boldsymbol{f}_i$ and historical journey as $\mathcal{J}_i = \{\boldsymbol{o}_t^i\}_{t=1}^{T_i} = \{{c}_t^i, \boldsymbol{q}_t^i, {g}_t^i, {w}_t^i\}_{t=1}^{T_i}$, where $\boldsymbol{o}_t^i$ is the observation of ad exposure at time step $t$, including advertising channel ${c}_t^i$, touch point feature vector $\boldsymbol{q}_t^i$, touch point gain ${g}_t^i$ (e.g., click, conversion), and cost ${w}_t^i$. 
% $T_i$ is the number of $u_i$'s touch points.

% Given a set of user $\mathcal{U}=\{u_1,u_2,\cdots,u_{\vert\mathcal{U}\vert}\}$ and a set of channels $\mathcal{C}=\{{c}_1,{c}_2,\cdots,{c}_{\vert\mathcal{C}\vert}\}$. Denote $u_i$'s static feature as $\boldsymbol{f}_i$ and historical journey as $\mathcal{J}_i = \{{c}_j^i, \boldsymbol{q}_j^i, {g}_j^i\}_{j=1}^{T_i}$, where ${c}_j^i$ is the advertising channel, $\boldsymbol{q}_j^i$ is the touch point feature vector, ${g}_j^i$ is the gain (e.g., click, conversion), at time step $j$. $T_i$ is the number of $u_i$'s touch points.

{\textbf{Channel Recommendation.}}
Supposing $\mathcal{C}$ is fixed, online service providers (e.g., advertising platforms) are interested in making an automated strategy for the advertiser that recommends a personalized channel $c_t^i\in\mathcal{C}$ for user $u_i$ based on her historical journey $\mathcal{J}_i$.  
In addition, the strategy should adjust based on new observations $\boldsymbol{o}_t^i$ after each ad exposure to meet the user's dynamic preferences. The goal is to select the most suitable channel to maximize each user's total gain (e.g., clicks, conversions), thereby boosting revenue. 

{\textbf{Budget Allocation.}}
In practice, online advertisers typically have a limited budget for advertising. Therefore, online service providers cannot mindlessly advertise but must constrain the channel recommendation and accurately filter out non-target customers to control cost risks, corresponding to channel-level and user-level budget allocation modules in Section~\ref{sec:budget}.

\subsection{Multi-task Offline Reinforcement Learning}
\label{sec:MDP}
\subsubsection{\textbf{Markov Decision Process}}
The problem can be formulated as a MDP, which consists of $(\mathcal{S},\mathcal{A},P,r)$. $\mathcal{S}$ is the state space, $\mathcal{A}$ is the action space, $P(s'\vert s,a)$ represents the transition dynamics, $r(s,a)$ is the reward function. In the user sequence, we denote $s_t, a_t, r_t$ as state, action, and reward at time step $t$. 
The policy $\pi$ can be deterministic or stochastic, which outputs a unique $a$ or $P(a\vert s)$ given state $s$, respectively. In our problem, $s$ represents the user state, $a$ means selecting ad channels for the user, and $r$ is the gain.

\subsubsection{\textbf{Optimization Objective.}}
When considering the limit, the MDP problem converts to the CMDP problem~\cite{liao2022cross,achiam2017constrained}, which is to maximize the expected cumulative rewards while constraining cumulative costs. 
In the advertising domain, the user interests are heterogeneous, and the corresponding click-through rate (CTR) and conversion rate (CVR) are different, so we consider the user dimension additionally.
We define a new optimization target that is formed as
\begin{equation}
\label{eq:CMDP}
   \max_{\pi,\pi_{u}} J_r(\pi,\pi_{u}) := \mathbb{E}\bigg[\sum_{t=1}^{T}\sum_{u\in\mathcal{U}_t}r_{t,u}\bigg], 
    J_w(\pi_{u}) = \mathbb{E}\bigg[\sum_{t=1}^{T}\sum_{u\in\mathcal{U}_t} w_{t,u}]\le W,
\end{equation}
where $\pi$ is the policy, $W$ is the budget of the advertiser. $\pi_{u} =\{\mathcal{U}_1,\cdots,\mathcal{U}_t,\cdots\}$ is a filter that selects the target users $\mathcal{U}_t\subset\mathcal{U}$ for advertising at $t$, and $r_{t,u},w_{t,u}$ are the reward and cost of ad exposure for user $u$. To solve the problem smoothly, we penalize the reward.
Specifically, we transform the problem into an unconstrained one by Lagrangian multiplier~\cite{zhang2021bcorle,tessler2018reward,achiam2017constrained}:
\begin{equation}
       L(\pi,\pi_{u},s) = J_r(\pi,\pi_{u})-s\cdot J_w(\pi_{u})=\mathbb{E}\bigg[\sum_{t=1}^{T}\sum_{u\in\mathcal{U}_t}r_{t,u}^*\bigg]+s\cdot W,
\end{equation}
where $r_{t,u}^* = r_{t,u} -s\cdot w_{t,u}$ is defined as the penalized reward, controlled by the penalty strength $s$. Therefore, our target is simplified to maximize the cumulative penalized reward. We penalize the reward, and for simplicity, we still name it the \textit{reward} in this paper.

\subsubsection{\textbf{Problem Statement.}}
We formally describe our tasks as: 
We learn an optimal policy $\hat{\pi}(\boldsymbol{a}\vert\boldsymbol{s})$ and reward prediction $\hat{r}(\boldsymbol{s})$ simultaneously, which means the model inputs a state $\boldsymbol{s}_t$, and then outputs the action prediction $\hat{\boldsymbol{a}}_t$ and reward prediction $\hat{\boldsymbol{r}}_t$. The prediction $\hat{\boldsymbol{a}}_t, \hat{\boldsymbol{r}}_t$ will be leveraged for channel recommendation policy (i.e., $\pi$) and user filter (i.e., $\pi_u$), respectively, introduced in Section~\ref{sec:Adv}.

% The objective is learning a policy that maximizes the expected cumulative rewards $\mathbf{E}[\sum_{t=1}^T r_t]$. In contrast to online RL, which constantly explores and exploits the environment, offline RL only utilizes collected datasets instead of further interacting with the environment. The challenge is learning the optimal policy based on user behavioral distribution without additional environmental interaction and observation collection.

% We formally describe our task as follows: 
% In the multi-task offline RL, we learn a policy $\hat{\pi}(\boldsymbol{a}\vert\boldsymbol{s})$ and reward prediction $\hat{r}(\boldsymbol{s})$ simultaneously, which means if we input a state $\boldsymbol{s}_t$, the model will output the action prediction $\hat{\boldsymbol{a}}_t$ and reward prediction $\hat{\boldsymbol{r}}_t$ at once. The prediction $\hat{\boldsymbol{a}}_t, \hat{\boldsymbol{r}}_t$ will be leveraged for channel selection and budget allocation,  respectively, introduced in Section~\ref{sec:Budget}.

% \begin{figure*}
%     \centering
%     \includegraphics[width=1.0\linewidth]{Architecture.pdf}
%     % \caption{Architecture}
%     \caption{The architecture of the proposed MTORL framework.}
%     \label{fig:Architecture}
% \end{figure*}

\section{Methodology}
\label{Sec:Methodology}

% As illustrated in Figure~\ref{fig:Architecture}, MTORL consists of five major components: causal state encoder, causal attention module, action and reward encoders, and optimization. We detail each component as follows.

As illustrated in Figure~\ref{fig:Architecture}, MTORL consists of five major components: the causal state encoder modeling sequential dependencies, the causal attention module enhancing user preferences modeling capabilities, action and reward encoders to predict corresponding actions and rewards for the agent, respectively, and the multi-task policy optimization. In the following, we detail each component.

\subsection{Embedding Module}
\label{sec:embedding}
Considering the conceptual discrepancy between advertising and offline RL, building a bridge to connect them is necessary. 
We propose a vital embedding module to map user historical exposure features into compact embedding vectors—representing states, actions, and rewards in MDP~\cite{arulkumaran2017deep,henderson2018deep}. Employing sequence modeling, we generate user sequence embeddings, which lay the foundation of the following encoder-decoder framework.

% To better conduct the offline RL in advertising and integrate users' time-involving preferences, we propose the embedding module to map user historical exposure features (e.g., user profiles, channel IDs, and user feedback) into dense embedding vectors (i.e., states, actions, and rewards) in MDP~\cite{arulkumaran2017deep,henderson2018deep}, then generate user sequence embeddings by sequence modeling.
 
\subsubsection{\textbf{Markov Decision Process}}
Specifically, for a user $u_i$, the state of $u_i$ at time step $t$ consists of user dynamic feature $\{{c}_t^i, \boldsymbol{q}_t^i, {g}_t^i, {w}_t^i\}$ and user static feature $\boldsymbol{f}_i$. We generate states, actions, and rewards (i.e., $\boldsymbol{s}_t$, $\boldsymbol{a}_t$, and $\boldsymbol{r}_t$) embeddings at each time step $t$ by
\begin{equation}
\label{eq: emb1}
\begin{aligned}
    &\textbf{State: }\boldsymbol{s}_t =\text{Concat}(\boldsymbol{q}_t, \boldsymbol{f}),\\
    &\textbf{Action: }\boldsymbol{a}_t = \mathrm{OnehotEncoder}({c}_{t}), \\
    & \textbf{Reward: }\boldsymbol{r}_t = \mathrm{MinMaxNorm}({g}_{t}-s{w}_{t}).
\end{aligned}
\end{equation}
$\mathrm{OnehotEncoder}(\cdot)$ (e.g., dummy variable) and $\mathrm{MinMaxNorm}(\cdot)$ (e.g., min-max normalization) are embedding functions generating action and reward, respectively. We default penalty strength $s$ in Equation~\eqref{eq: emb1} as a constant $0.5$ for consistency~\cite{tessler2018reward,zhang2021bcorle}.
We will detail our implementations under different settings in Appendix~\ref{a1}.

% $\mathrm{OnehotEncoder}(\cdot)$ (e.g., dummy variable~\cite{hardy1993regression}) and $\mathrm{MinMaxNorm}(\cdot)$ (e.g., min-max normalization~\cite{patro2015normalization}) are embedding functions generating action and reward, respectively. We default penalty strength $s$ in Equation~\eqref{eq: emb1} as a constant $0.5$ for consistency~\cite{tessler2018reward,zhang2021bcorle}.
% We will detail our implementations under different settings in Appendix~\ref{a1}.

\subsubsection{\textbf{Sequence Modeling}}
\label{subsubsec: sequence modeling}
We consequently merge the embeddings to model users' sequential behaviors (i.e., generate a sequence embedding $\boldsymbol{X}$) as follows
\begin{equation}
\label{eq: embedding}
   \boldsymbol{X} = \{\boldsymbol{x}_t\}_{t=1}^n = \{\boldsymbol{W}_e\cdot\text{Concat}(\boldsymbol{a}_{t-1}, \boldsymbol{r}_{t-1}, \boldsymbol{s}_t)\}_{t=1}^n,
\end{equation}
where $\boldsymbol{W}_e\in\mathbb{R}^{d\times F}$ is the embedding matrix, $F$ is the feature size, $n$ is the sequence length. 
To study how different lengths impact learning performance, we conduct comprehensive experiments in Section~\ref{sec:length}, representing an interesting trade-off between the quality and quantity of captured preferences: a moderate $n$ could introduce more reliable sequential dependencies and less noise.

% We consequently merge the embeddings to model users' sequential behaviors (i.e., generate a sequence embedding $\boldsymbol{X}$) and learn the behavioral policy (i.e., the probability of user $u$ select action $\boldsymbol{a}$ at time step $t$, $\pi(\boldsymbol{a}_t\vert \boldsymbol{x}_{t-n+1:t})$, where $n$ is the sequence length) as follows
% \begin{equation}
% \label{eq: embedding}
%    \boldsymbol{X} = \{\boldsymbol{x}_t\}_{t=1}^n = \{\boldsymbol{W}_e\cdot\text{Concat}(\boldsymbol{a}_{t-1}, \boldsymbol{r}_{t-1}, \boldsymbol{s}_t)\}_{t=1}^n,
% \end{equation}
% where $\boldsymbol{W}_e\in\mathbb{R}^{d\times F}$ is the embedding matrix, $F$ is the feature size. Due to users' current preferences being mainly influenced by their recent historical exposures, we do not take a large length $n$ ($n\le 30$)~\cite{ji2017additional,ren2018learning}. To study how different lengths impact learning performance, we conduct comprehensive experiments in Section~\ref{sec:length}, representing an interesting trade-off between the quality and quantity of captured preferences: a smaller $n$ could introduce more reliable sequential dependencies and less noise into RL.

\subsection{Causal State Encoder}
In advertising, the return (e.g., conversion) is influenced by current exposure and its connection to a series of preceding user exposure events. The challenge is integrating the causality between previous and current states into sequence embeddings to simulate users' dynamic interests. 
Towards this end, we propose the causal state encoder, for each user $u$, aggregating her previous behaviors and current state by the causal temporal convolutional network (TCN)~\cite{bai2018empirical}. The causal TCN leverages the dilated causal convolution to capture temporal dependencies in the sequence.
Mathematically, we generate the hidden layer $\boldsymbol{H}^{(l)} = (\boldsymbol{h}_1^{(l)},\cdots,\boldsymbol{h}_n^{(l)})$ as
\begin{equation}
\label{eq:causalconv}
       \boldsymbol{H}^{(l)}= \mathrm{DilatedCausalConv}(\boldsymbol{H}^{(l-1)})
        \: ;\:
       l=1,2,\cdots,L_1, 
\end{equation}
% \begin{equation}
%     \begin{aligned}
%        (\boldsymbol{h}_1^{(l)},\cdots,\boldsymbol{h}_n^{(l)})&= \mathrm{DilatedCausalConv}(\boldsymbol{h}_{1}^{(l-1)},\cdots,h_n^{(l-1)}),\\
%        (\boldsymbol{h}_1^{(0)},\cdots,\boldsymbol{h}_n^{(0)})&= (\boldsymbol{x}_1,\cdots,\boldsymbol{x}_n)
%        \: ;\:
%        l=1,2,\cdots,L_1,      
%     \end{aligned}
% \end{equation}
where $\boldsymbol{H}^{(0)}= \boldsymbol{X}$ is the input sequence embeddings. The causal convolution for generating the hidden state is formulated as $\boldsymbol{h}_t^{(l)} = \sum_{i=0}^{k_c-1}f(i)\cdot \boldsymbol{h}_{t-d_c\cdot i}^{(l-1)}$,
% \begin{equation}
%     \boldsymbol{h}_t^{(l)} = \sum_{i=0}^{k_c-1}f(i)\cdot \boldsymbol{h}_{t-d_c\cdot i}^{(l-1)},
% \end{equation}
where $d_c,k_c$ represents the dilation factor and filter size of causal convolution, and $f$ is the filter function. 
According to the formulation, the hidden state at time $t$ only depends on the past hidden state, i.e., the hidden state at time $k (k\le t)$, of the previous layer instead of any future information, providing the causality.
After that, we add residual connections~\cite{he2016deep} and other layers (e.g., Dropout and Normalization~\cite{salimans2016weight}) to enhance the sequence representation learning. Finally, we 
make a linear transformation for the causal TCN's output $\boldsymbol{H}^{(L_1)}$ to get the causal state sequence:
\begin{equation}
   \label{eq:causal state encoder}
   \tilde{\boldsymbol{S}} = \mathrm{LeakyReLU}(\boldsymbol{W}_s\boldsymbol{H}^{(L_1)}+\boldsymbol{B}_s),
\end{equation}
where $\boldsymbol{W}_{s}\in\mathbb{R}^{d\times d}, \boldsymbol{B}_{s}\in\mathbb{R}^{d\times n}$ are the weight and bias, respectively. In particular, $\tilde{\boldsymbol{s}}_t$ represents the causal state at time step $t$, LeakyReLU~\cite{maas2013rectifier} is a variant of ReLU activation.

\subsection{Causal Attention Module}
In advertising, user historical behaviors showcase different impacts of user current preferences, e.g., meaningless accidental exposures or repeated behavior patterns~\cite{chen2022denoising,qin2021world}. To better reflect user preferences, we introduce the causal attention mechanism, which is the foremost part of the GPT~\cite{radford2018improving}. Different from the standard self-attention mechanism~\cite{vaswani2017attention}, causal attention uses a causal mask on self-attention to guarantee auto-regressive modeling and cut off information from future time steps. Formally, the causal attention module takes the causal state sequence $\tilde{\boldsymbol{S}}\in\mathbb{R}^{d\times n}$ as input and outputs causal attention scores $\boldsymbol{A}$ as follows
\begin{equation}
\label{eq: causal attention module}
    \boldsymbol{A}=\mathrm{CausalAttn}(\tilde{\boldsymbol{S}}^\mathrm{T}\boldsymbol{W}_Q,\tilde{\boldsymbol{S}}^\mathrm{T}\boldsymbol{W}_K,\tilde{\boldsymbol{S}}^\mathrm{T}\boldsymbol{W}_V),
\end{equation}
where $\boldsymbol{W}_Q, \boldsymbol{W}_K, \boldsymbol{W}_V\in\mathbb{R}^{d\times d}$ are projection matrices. To alleviate overfitting issues and accelerate the training process~\cite{liu2023linrec}, we introduce the residual connection~\cite{he2016deep} and layer normalization~\cite{ba2016layer} as follows:
\begin{equation}
\begin{aligned}
    \boldsymbol{M}^{(l)}&=\mathrm{LayerNorm}(\boldsymbol{M}^{(l-1)}+\mathrm{Dropout}(\boldsymbol{S}^{(l-1)})),\\ 
    \boldsymbol{S}^{(l-1)}&=\mathrm{LeakyReLU}(\boldsymbol{W}_M^{(l-1)}\boldsymbol{M}^{(l-1)} + \boldsymbol{B}_M^{(l-1)}),\\
    \boldsymbol{M}^{(0)}&=\boldsymbol{A}^\mathrm{T}
    \: ;\:
    l = 1, 2, \cdots, L_2, 
 \end{aligned}
\end{equation}
where $\boldsymbol{W}_M^{(l)}\in\mathbb{R}^{d\times d}, \boldsymbol{B}_M^{(l)}\in\mathbb{R}^{d\times n}$ are the weight and bias of layer $l$. 
We can learn behavior-aware sequential dependencies based on the above modules (i.e., causal state encoder and causal attention module). To further accommodate advertising, we devise two decoders (i.e., action decoder and reward decoder) based on the encoder-decoder architecture in the following.

\subsection{Action Decoder}
The channel recommendation problem entails discerning the user's optimal course of action (i.e., channel). To address this, we employ an action decoder designed to extract and interpret acquired prior information from preceding modules. This process facilitates the generation of a refined action policy, thereby enhancing the precision and effectiveness of the decision-making mechanism.
With the aforementioned casual attention module, we can calculate the probability $\hat{y}_{tj}$ that a user chooses an action $j$ at time step $t$ by
\begin{equation}
\label{eq: action decoder}
\begin{aligned}
    \boldsymbol{Z} &= (\boldsymbol{W}_Z\boldsymbol{M}^{(L_2)} + \boldsymbol{B}_Z)^\mathrm{T},\\
    \hat{y}_{tj}&=\frac{\exp(z_{tj})}{\sum_{j=1}^m\exp(z_{tj})},
\end{aligned}
\end{equation}
where $\boldsymbol{W}_Z\in\mathbb{R}^{m\times d}$  linearly transforms representations from hidden space into action space, $\boldsymbol{B}_Z\in\mathbb{R}^{m\times n}$ is bias, and $z_{tj}$ is the action score at time step $t$ of action $j$.
For the implementation, such a design choice allows us to suit different settings to make action decisions. For example, if we apply a deterministic policy, the predicted action $\hat{\boldsymbol{a}}_{t}$ at time step $t$ should be the action with the highest $\hat{y}_{tj}$. Therefore, we can formulate the deterministic policy by
\begin{equation}
\label{eq:policy_deterministic}
    \hat{\pi}(\boldsymbol{a}\vert \boldsymbol{x}_{t-n+1:t}) = \mathop{\arg\max}_{\boldsymbol{a}^j\in\{\boldsymbol{a}^1,\cdots,\boldsymbol{a}^m\}}{\hat{y}_{tj}}.
\end{equation}
Moreover, we can seamlessly adopt the stochastic policy as follows
\begin{equation}
\label{eq:policy_stochastic}
    \hat{\pi}(\boldsymbol{a}\vert\boldsymbol{x}_{t-n+1:t}) = \boldsymbol{a}^j\quad\text{with probability }\hat{y}_{tj}. 
\end{equation}
% Therefore, for the training process, we output all times predicted actions $\{\hat{\boldsymbol{a}}_{t}\}_{t=1}^n$; for the inference process, we output the last time predicted action ${\hat{a}_{n}}$. 
Therefore, during the training process, we can predict actions (i.e., $\hat{\boldsymbol{a}}_{t}$) at each time step and force the agent to simulate users' sequential behaviors for subsequent channel recommendation inferences.

\subsection{Reward Decoder}
Predicting rewards is more challenging due to the offline RL settings. In practical scenarios, we have to learn rewards based on actions to optimize the agent, but we cannot obtain the ground truth (i.e., $\boldsymbol{a}_t$). To tackle the issues, previous offline RL studies~\cite{ho2016generative,finn2016guided} used predicted actions (i.e., $\hat{\boldsymbol{a}}_{t}$) to further predict rewards. However, such an approach is unsuitable for recent advantages (i.e., leveraging the causal attention module to encode states). Theoretically, the attention mechanism is \textit{not} Lipschitz continuous~\cite{kim2021lipschitz}, meaning that potential noise in the inputted causal state may significantly change the attention's output of action predictions as the layer deepens.  

% Unlike the action encoder, predicting rewards is more challenging due to the offline RL settings. In practical scenarios, we have to learn rewards based on actions to optimize the agent, but we cannot obtain the ground truth (i.e., $\boldsymbol{a}_t$) of the user's action at time step $t$. To tackle the issues, previous offline RL studies~\cite{ho2016generative,finn2016guided} used predicted actions (e.g., $\hat{\boldsymbol{a}}_{t}$) to learn rewards. However, such an approach is unsuitable for recent advantages (i.e., leveraging causal attention models to encode states). Theoretically, the attention mechanism is \textit{not} Lipschitz continuous~\cite{kim2021lipschitz}, meaning that potential noise in the inputted causal state may significantly change attention's output of action predictions as the layer deepens. Some studies leverage regularization methods (e.g., Jacobian regularization~\cite{chen2022denoising}) to devise a more robust attention architecture. However, additional computational costs make these methods impractical in advertising scenarios. 

To prevent cumulative errors, we directly leverage the causal states $\tilde{\boldsymbol{s}}_t$ instead of $\hat{\boldsymbol{a}}_{t}$ to predict reward, retaining the flexibility of the auxiliary model meanwhile exploiting the learned prior information. The logic behind our design choice is that the proposed causal attention module can be treated as an ``information filter'', emphasizing highly related user exposures by assigning larger attention weights. Therefore, leveraging causal states (i.e., $\{\tilde{\boldsymbol{s}}_t\}_{t=1}^n$) to predict rewards can inherently prevent the overestimated rewards caused by the exaggeration of the attention mechanism for some noisy patterns.
Specifically, we formulate the reward decoder to predict rewards (i.e., $\hat{\boldsymbol{r}}$) as follows:
\begin{equation}
\label{eq: reward decoder}
\begin{aligned}
\boldsymbol{N}^{(l)}&=\mathrm{LeakyReLU}(\boldsymbol{W}_N^{(l-1)}\boldsymbol{N}^{(l-1)} + \boldsymbol{b}_N^{(l-1)}),\\
\boldsymbol{N}^{(0)}&=\tilde{\boldsymbol{S}}
    \: ;\:
l = 1, 2, \cdots, L_3, \\
\hat{\boldsymbol{r}} &= f(\boldsymbol{W}_r\boldsymbol{N}^{(L_3)} + \boldsymbol{B}_r),  
\end{aligned}
\end{equation}
where $\boldsymbol{W}_N^{(l)},\boldsymbol{B}_N^{(l)}$ are the weight and bias of layer $l$. $\boldsymbol{W}_r,\boldsymbol{b}_r$ are weight and bias at the output layer, $f(\cdot)$ is an activation function. 
%%%%%%5
$f(\cdot)$ is an activation function (e.g., ReLU, Sigmoid) to obtain the final reward prediction. 
We detail $f(\cdot)$ under different settings (e.g., binary, continuous, or multi-class rewards) in Appendix~\ref{a2}.

The insight here is that using the predicted action $\hat{\boldsymbol{a}}_{t}$ to further predict rewards is a two-hop approach, which may bring cumulative errors. So, we predict reward based on causal states, which are obtained in a shallower layer, implicitly leveraging prior information of action prediction while mitigating the risk of error cumulation.
 
% To prevent cumulative errors, we directly leverage the causal states $\tilde{\boldsymbol{s}}_t$ instead of $\hat{\boldsymbol{a}}_{t}$ to predict reward, retaining the flexibility of the auxiliary model meanwhile exploiting the prior information learned from the main model. The logic behind our design choice is that the proposed casual attention module can be treated as an ``information filter'', emphasizing highly related user exposures by assigning larger attention weights. Therefore, leveraging causal states (i.e., $\{\tilde{\boldsymbol{s}}_t\}_{t=1}^n$) to predict rewards can inherently prevent the overestimated rewards caused by the exaggeration of the attention mechanism for some noisy patterns.
% Specifically, we formulate the reward decoder to predict rewards (i.e., $\hat{\boldsymbol{r}}$) as follows:
% \begin{equation}
% \label{eq: reward decoder}
% \begin{aligned}
% \boldsymbol{N}^{(l)}&=\mathrm{LeakyReLU}(\boldsymbol{W}_N^{(l-1)}\boldsymbol{N}^{(l-1)} + \boldsymbol{b}_N^{(l-1)}),\\
% \boldsymbol{N}^{(0)}&=\tilde{\boldsymbol{S}}
%     \: ;\:
% l = 1, 2, \cdots, L_3, \\
% \hat{\boldsymbol{r}} &= f(\boldsymbol{W}_r\boldsymbol{N}^{(L_3)} + \boldsymbol{B}_r),  
% \end{aligned}
% \end{equation}
% where $\boldsymbol{W}_N^{(l)}\in\mathbb{R}^{d\times d}$ and $,\boldsymbol{B}_N^{(l)}\in\mathbb{R}^{d\times n}$ are the weight and bias of layer $l$. $\boldsymbol{W}_r$ and $\boldsymbol{b}_r$ are weight and bias at the output layer, $f(\cdot)$ is an activation function. 

\subsection{Multi-task Policy Optimization}
\label{sec:opt}

Then, the critical challenge is to improve predictions of action and reward decoders jointly. In addition, it is necessary to learn the policy with high cumulative rewards. In order to solve these problems together, we propose a multi-task optimization objective that comprises three components: a primary focus on policy learning and two auxiliary parts—reward learning and DPO.

\noindent\textbf{Policy Learning Optimization.} Specifically, we formulate the policy learning task as cross-entropy (CE) loss by minimizing the information entropy between predicted actions and ground truth:
\begin{equation}
\label{eq:CE}
    \mathcal{L}_{Policy} = -\mathbb{E}_{\boldsymbol{x}=x_n,x_{n-1}\cdots,x_1\in\mathcal{D}}\Big[\frac{1}{n}\sum_{t=1}^n\sum_{j=1}^m y_{tj} \log(\hat{y}_{tj})\Big],
\end{equation}
where $\boldsymbol{x}$ are continuous exposures from the user journeys $\mathcal{D}$.

\noindent\textbf{Reward Learning Optimization.} For binary rewards (e.g., $r_t = \{0,1\}$), we define the auxiliary task of predicting reward as
\begin{equation}
\label{eq: L2}
    \mathcal{L}_{Reward} = -\mathbb{E}_{\boldsymbol{x}\in\mathcal{D}}\Big[\frac{1}{n}\sum_{t=1}^n \boldsymbol{r}_{t}\log(\hat{\boldsymbol{r}}_{t}) + (1- \boldsymbol{r}_{t})(1-\log(\hat{\boldsymbol{r}}_{t}))\Big].
\end{equation}
Note that for other settings (e.g., multi-class or continuous rewards), we also formulate variants of Eq.~(\ref{eq: L2}) in Appendix~\ref{a3}.

\noindent\textbf{Direct Preference Optimization Loss.} 
To directly optimize model preferences, i.e., maximize the cumulative reward, we design a DPO loss~\cite{rafailov2024direct} for the proposed model, formulated as
\begin{equation}
\label{eq:DPO}
    \mathcal{L}_{DPO} = -\mathbb{E}_{\boldsymbol{x}^w,\boldsymbol{x}^l\in\mathcal{D}}\Big[\frac{1}{n}\sum_{t=1}^n \log\sigma(\beta\log\big(\pi(a^w_t\vert \boldsymbol{x}^w_t)-\beta\log(\pi(a^l_t\vert \boldsymbol{x}^l_t)\big)\Big].
\end{equation}
where $\{x^w,x^l\}$ is a pair of user sequences sampled from the dataset $\mathcal{D}$ with total rewards $\sum_{t=1}^n \boldsymbol{r}^w_{t}>\sum_{t=1}^n \boldsymbol{r}^l_{t}$, and we use the default value $\beta=0.1$. 
% We simplify the original DPO loss by omitting the KL divergence since we don't use a reference model here. 
% In addition, we consider a pair of sequences instead of a pair of points, as we focus on long-term returns.
Therefore, minimizing DPO loss makes the model prefer to apply a policy that brings higher long-term returns. 

% \noindent\textbf{Online Exploration Optimization.} Moreover, to reinforce the action prediction and make it suitable for online exploration, we add a Shannon-Entropy (SE) loss into the main loss as
% \begin{equation}
%     \mathcal{L}_{SE} = \mathbb{E}_{\boldsymbol{x}\in\mathcal{D}}[\frac{1}{n}\sum_{t=1}^n\sum_{j=1}^m \hat{y}_{tj} \log_2(\hat{y}_{tj})].
% \end{equation}

\noindent\textbf{Final Optimization Objective.} According to the above three objectives, we can formulate our final optimization objective by
\begin{equation}
\label{eq:loss}
    \mathcal{L} = \mathcal{L}_{Policy} + \mu\mathcal{L}_{Reward} + \lambda\mathcal{L}_{DPO},
\end{equation}
where $\mu, \lambda$ are the tuning parameters, adjusting contributions of each loss function for the multi-task policy optimization. To verify the impact of different parameter values, we carefully tune the parameters (i.e., $\mu$ and $\lambda$) in Section~\ref{sec:Parameter}, which shows small values (e.g., $0.08$ and $1.4$) can affect the final performance, demonstrating the effectiveness of our multi-task policy optimization.

\subsection{Online Advertising}
\label{sec:Adv}

Based on the above components, we can pre-train an \name model on the offline dataset, which can well simulate user behaviors. This section will technically detail how to deploy \name for online advertising. 
% The pseudo-codes are specified in Appendix~\ref{appendix:algo}. 
The pseudo-codes are specified in Algorithm~\ref{algorithm}. 

\begin{algorithm}[t]
	\caption{Online advertising procedure}
	\label{algorithm}
	\begin{algorithmic}[1]
		\STATE\textbf{Input:} A trained \name model, initial policy $\mathcal{P}_{initial}$, budget $W$, channel cost $\{w_j\}_{j=1}^m$, factor $\eta$,  total rounds $K$ 
		\STATE\textbf{Exploration:} Initialize journey memory $\mathcal{J}$ by $\mathcal{P}_{initial}$
        \STATE $\mathcal{P},\mathcal{R}^s\leftarrow \name(\mathcal{J})$ \quad// Initialize policy and reward 
        \STATE\textbf{Exploitation:}\WHILE {$k\le K$ and $W>0$}
        \STATE Rank users $\mathcal{U}$ using $\mathcal{R}$ and select top-$N$ user 
        % \STATE Rank users $\mathcal{U}$ using $\mathcal{R}^s$ and select top-$N$ user
        % $\{u_1,\cdots, u_N\}$
            \FOR{$i=1,\cdots, N$}
            \STATE Use policy $\mathcal{P}_{u_i}$ to recommend a channel $c_{u_i}$
            \STATE Advertising for $u_i$ from channel $c_{u_i}$ and observe $\boldsymbol{o}_{u_i}$
            \STATE Update $u_i$'s memory $\mathcal{J}_{u_i}$ by new observation $\boldsymbol{o}_{u_i}$  
            \STATE $\mathcal{P}_{u_i},\mathcal{R}_{u_i}\leftarrow \name(\mathcal{J}_{u_i})$
            \quad// predict action and reward
            % \STATE $\mathcal{R}^s_{u_i}\leftarrow \mathcal{R}_{u_i}-s\cdot w_{u_i}$ \quad// calculate penalized reward
            \STATE $\mathcal{P}_{u_i}\leftarrow \eta\mathcal{P}_{u_i}+(1-\eta)\mathcal{P}_{initial}$
            \quad// weight averaging 
            \STATE $W\leftarrow W-w_{u_i}, k\leftarrow k+1$\quad// reduce the budget
            \ENDFOR
        \ENDWHILE
	\end{algorithmic}  
\end{algorithm}

\subsubsection{\textbf{Channel Recommendation}}
\label{sec:channel}
We need to recommend an appropriate channel for each user to maximize business revenue. Specifically, for user $u_i$, we input her recent journey memory $\mathcal{J}_{u_i}$ into \name, as shown in Figure~\ref{fig:Architecture}(b), and infer 
the policy $\mathcal{P}_{u_i}$, which is determined by the predicted action (Eq. \eqref{eq:policy_deterministic}). The action prediction is specified in Figure~\ref{fig:Architecture}(a), and notice that only the last prediction is used for inference, different from the training process. Then, we recommend channel $c_{u_i}$ for $u_i$ based on policy $\mathcal{P}_{u_i}$. 

% Recall that we have to recommend an appropriate channel for each user to maximize business revenue. Therefore, for each user, we input her historical data into the proposed \name, as shown in Figure~\ref{fig:Architecture}(a), and generate corresponding action predictions by the proposed action decoder (i.e., Eq.~(\ref{eq: action decoder})). Specifically, we generate base embeddings (i.e., $\boldsymbol{x}$) by the embedding module (i.e., Eq.~(\ref{eq: emb1})-(\ref{eq: embedding})) and input them to the causal state encoder (i.e., Eq.~(\ref{eq:causal state encoder}) and causal attention module (i.e., Eq.~(\ref{eq: causal attention module}) to encode user historical behaviors so as to predict actions (recommend channels) according to the action encoder (i.e., Eq.~(\ref{eq: action decoder})).

\subsubsection{\textbf{Budget Allocation}}
\label{sec:budget}
It is crucial to adjust the weights of different channels from a global view and dynamically filter target users from the entire user set simultaneously to maximize the gains under the budget constraint. To address these two challenges, we propose a budget allocation strategy from both channel- and user-level, as shown in Figure~\ref{fig:Architecture}(b).

% We have to dynamically filter target users $\mathcal{U}_t$ from the entire user set and adjust the weights of different channel from a global view simultaneously to maximize the gains under the budget constraint. To address these two challenges, we subsequently propose a hierarchical budget allocation strategy from both channel- and user-side, as shown in Figure~\ref{fig:Architecture}(b).

% Unlike the channel recommendation problem, we have to filter users from the entire user set to maximize the average gains of user-channel recommendations under budget constraints. To comprehensively model user preferences, we subsequently propose a hierarchical budget allocation strategy from both channel- and user-side, as shown in Figure~\ref{fig:Architecture}(b). Specifically, We propose a channel-level budget allocation module to evaluate the potential benefits of advertising on different channels. In addition, we devise a user-level budget allocation module to determine the reliability of a target user, thus discriminating whether advertising to user meets expected revenues.

\noindent{\textbf{Channel-level Allocation.}}
In practice, the importance of channels to gains (e.g., click) is different~\cite{ji2017additional,ren2018learning}. 
Learning such prior information from offline datasets and exploiting it in online advertising is challenging. 
To tackle this problem, we propose to learn both explicit and implicit policies.
Specifically, we straightly leverage the CTR~\cite{kingsnorth2022digital} to represent channels' importance so as to learn the explicit policy. Accordingly, we calculate the explicit policy by $\mathcal{P} = [p_1, p_2,\cdots,p_m]$, where $p_j = CTR_j/\sum_{i=1}^m CTR_i$ and $CTR_j$ is the CTR of channel $j$. Despite being effective, ubiquitous sparse datasets may lead to bias issues (e.g., popularity bias~\cite{abdollahpouri2019unfairness}), making the explicit policy exaggerate some channels' importance. To overcome this challenge, we learn the implicit policy by filtering out unreliable conversions. Technically, we leverage the predicted rewards (i.e., Eq.~(\ref{eq: reward decoder})) serving as the threshold, which is commonly used in transfer learning scenarios (e.g., cross-domain recommendations~\cite{tang2012cross}) to adapt unbalanced data distributions. Therefore, we formulate the implicit budget ratio as $\hat{\mathcal{P}} = [\hat{p}_1, \hat{p}_2,\cdots,\hat{p}_m]$, where $p_j = \hat{CTR}_j/\sum_{i=1}^m \hat{CTR}_i$, $\hat{CTR}_j = \sum\mathbb{I}(\hat{r}_t>\tau\vert a^j)/N_j$, $N_j$ is the channel $j$'s ad exposures and $\mathbb{I}(\cdot)$ is the indicator function filtering reliable conversions. Afterward, we merge the explicit and implicit ratios to obtain the initial policy:
\begin{equation}
\label{eq:behavior}
\mathcal{P}_{initial} = (1-\alpha)\mathcal{P} + \alpha\hat{\mathcal{P}},
\end{equation}
where $\alpha$ is a tuning parameter, and $\mathcal{P}_{initial}$ is leveraged in the initial exploration and adjusting the recommendation policy.

\noindent{\textbf{User-level Allocation.}}
We select the target users by sorting the user's reliability to reduce meaningless costs.
Like the generation of policies, we use \name to predict rewards $\mathcal{R}_{u_i}$ for each user $u_i$ according to her journey memory $\mathcal{J}_{u_i}$.
Then, we apply $\mathcal{R}$ as the sorting criterion of reliability.
%%%%%%%
Since the learned reward model can directly reflect users' conversion tendencies, which are highly related to CTR and GMV. Moreover, in a scenario with massive users and new streaming users, it is difficult to model the action space for selecting users to conduct RL (the potential space is too large and complicated). Therefore, we use the predicted value of $\mathcal{R}$ as a ranking indicator for users, which can more accurately and efficiently filter high-conversion-tendency users.
 
% We select the target users by sorting the user's reliability to reduce meaningless costs.
% Like the generation of policies, we use \name to predict rewards $\mathcal{R}_{u_i}$ for each user $u_i$ according to her journey memory. We can calculate the penalized rewards by $\mathcal{R}_{u_i}^{s} = \mathcal{R}_{u_i}-s\cdot w_{u_i}$, where $w_{u_i}$ is the cost of channel $c_{u_i}$. 
% Then, we apply $\mathcal{R}^s$ as the sorting criterion of reliability.

\subsubsection{\textbf{Advertising Procedure}}
\label{sec:adv_procedure}
Then, we design an automated advertising procedure integrating the above components. At the \textbf{Exploration} phase, we conduct advertising to all users $\mathcal{U}$ applying the stochastic policy $\mathcal{P}_{initial}$ to initialize journey memory $\mathcal{J} = \{\mathcal{J}_u, u\in\mathcal{U}\}$ for alleviating the cold-start problem~\cite{lika2014facing}. Next, we feed $\mathcal{J}$ into \name to generate policy memory $\mathcal{P}$ and reward memory $\mathcal{R}$. 
Then, we get into the \textbf{Exploitation} phase.
At each round, we use $\mathcal{R}^s$ to rank top-$N$ as this round's target users, and utilize $\mathcal{P}$ to recommend the advertising channels for them. We advertise to target users based on recommendations, get user feedback, and update their journey memories (by adding new observations and dropping early memories). Subsequently, we update their policy and reward memories using \name, prepared for the next round. The procedure will be terminal while the maximum budget or rounds is reached. 

In the real-time serving environments, we could store the streaming data (i.e., user journey memory $\mathcal{J}$) and leverage incremental training methods to update the model hourly or daily (e.g., mini-batch training or fine-tuning). This is efficient and stable for real-time serving, not requiring entire re-training.

\begin{table}[t]
% \vspace{-2mm}
\fontsize{8}{11}\selectfont
  \caption{Statistics of the datasets.}
  % \vspace{-2mm} 
  \begin{tabular}{ccc}
    \toprule
    Datasets & \# Users & \# Interactions \\
    \midrule
    \textbf{KuaiRand-Pure}  & $27,285$      & $1,436,609$ \\
    \textbf{Criteo}         & $6,142,256$   & $16,468,027$  \\
    % \textbf{Commercial}     & $8,626,849$   & $21,607,181$  \\
    \bottomrule
  \end{tabular}
  \label{tab:statistics}
   % \vspace{-3mm} 
\end{table}

\begin{table*}[t]
    \fontsize{8}{11}\selectfont
    \caption{\label{tab:fit}Overall accuracy performance comparison. 
    The best result is bold and the second-best result is underlined in each row. All improvements are \textbf{statistically significant} (i.e., two-sided t-test with $p<0.05$) over baselines.}
        % \vspace{-3mm} 
    {\begin{threeparttable}
    \begin{tabular}{cccccccccccccc}
        \toprule
        % \toprule
        \multirow{1}{*}{Datasets}&\multirow{1}{*}{Metrics}&
        W\&D & DIN & STAR & MIREC & R-BCQ & AG & BC & CQL & IQL & DT & ODT & \textbf{\name} \cr
        
        \cmidrule(lr){1-14}
        \multirow{3}{*}{KuaiRand-Pure}
        & F1-score  & 0.5210 & 0.5649 & 0.5731 & 0.5723 & 0.6148 & 0.6060 & 0.5862 & 0.5970 & 0.6203 & 0.6296 & \underline{0.6319} & \textbf{0.6793} \cr
        & Precision & 0.5464 & 0.5692 & 0.5744 & 0.5735 & 0.6154 & 0.6081 & 0.5863 & 0.5980 & 0.6251 & 0.6327 & \underline{0.6390} & \textbf{0.6811} \cr
        & Recall    & 0.5269 & 0.5669 & 0.5738 & 0.5727 & 0.6150 & 0.6065 & 0.5862 & 0.5972 & 0.6203 & 0.6308 & \underline{0.6335} & \textbf{0.6798} \cr
        
        \cmidrule(lr){1-14}
        \multirow{3}{*}{Criteo}
        & F1-score  & 0.2712 & 0.2974 & 0.2956 & 0.3034 & 0.3440 & 0.2982 & 0.2850 & 0.3395 & 0.3834 & \underline{0.4303} & 0.4168 & \textbf{0.5002} \cr
        & Precision & 0.2966 & 0.3118 & 0.3092 & 0.3170 & 0.3854 & 0.3239 & 0.3311 & 0.3906 & 0.4230 & \underline{0.4922} & 0.4738 & \textbf{0.5424} \cr
        & Recall    & 0.2663 & 0.2808 & 0.2878 & 0.2991 & 0.3326 & 0.2934 & 0.2659 & 0.3173 & 0.3655 & \underline{0.4101} & 0.3918 & \textbf{0.4893} \cr

        % \cmidrule(lr){1-14}
        % \multirow{3}{*}{Commercial}
        % & F1-score  & 0.2115 & 0.2247 & 0.2266 & 0.2309 & 0.2576 & 0.2231 & 0.2132 & 0.2542 & 0.2875 & \underline{0.3224} & 0.3122 & \textbf{0.3756} \cr
        % & Precision & 0.2276 & 0.2389 & 0.2370 & 0.2433 & 0.2795 & 0.2353 & 0.2406 & 0.2832 & 0.3072 & \underline{0.3570} & 0.3438 & \textbf{0.3939} \cr
        % & Recall    & 0.2034 & 0.2213 & 0.2239 & 0.2285 & 0.2457 & 0.2168 & 0.1966 & 0.2339 & 0.2699 & \underline{0.3032} & 0.2890 & \textbf{0.3615} \cr

        \bottomrule
        % \bottomrule
    \end{tabular}
    % \vspace{0mm}
    \end{threeparttable}}
        % \vspace{-3mm}
\end{table*}

\section{Experiments}
\label{sec:experiment}
In this section, we aim to answer the following research questions:
\begin{itemize}[leftmargin=*]
\item \textbf{RQ1}: How does \name perform compared with other methods regarding behavioral policy learning?
\item \textbf{RQ2}: How do different components contribute to \name?
\item \textbf{RQ3}: How does \name influenced by tuning parameter?
\item \textbf{RQ4}: How effectiveness of \name in online environment?
% \item \textbf{RQ4}: How is the effectiveness of \name's channel selection and budget allocation in the online environment?
% \item \textbf{RQ4}: How does \name perform in real e-commerce platform?
\end{itemize}

\subsection{Experimental Settings}

\subsubsection{\textbf{Datasets and Metrics}} 
To evaluate the effectiveness of \name, we conduct experiments on two benchmark datasets: (1) 
\textbf{KuaiRand-Pure\footnote{\url{https://kuairand.com/}}}: it contains users' unbiased journey data with random exposure. (2) \textbf{Criteo\footnote{\url{http://ailab.criteo.com/criteo-attribution-modeling-bidding-dataset/}}}: it contains Criteo live traffic data. Each sample represents one impression exposed to a user. 
% (3) \textbf{Commercial\footnote{We will publicize the dataset to foster research on this important topic.}}: We also collect a private dataset from our commercial short-video platform to evaluate MTORL in real-world applications.
For each user, we form her historical exposures chronologically as journey $\mathcal{J}_i$, specified in Section~\ref{sec:ad}. 
We continue to process datasets by extracting the user exposure sequences with the minimal length of $10$. 
In particular, to simulate multi-channel advertising, we regard the video types and campaign categories in three datasets as channels, respectively, following previous work~\cite{kumar2020camta,yao2022causalmta,ren2018learning}. The statistical information is provided in Table~\ref{tab:statistics}. More information on datasets is introduced in Appendix~\ref{sec:app_dataset}. We consider widely used metrics, \textit{F1-score}, \textit{Precision}, and \textit{Recall}.
% ~\cite{chowdhery2022palm,dosovitskiy2020image,he2020lightgcn}

% To evaluate the effectiveness of \name, we conduct experiments on two benchmark datasets: (1) 
% \textbf{KuaiRand-Pure\footnote{\url{https://kuairand.com/}}}: it contains users' unbiased journey data with random exposure. (2) \textbf{Criteo\footnote{\url{http://ailab.criteo.com/criteo-attribution-modeling-bidding-dataset/}}}: it contains Criteo live traffic data. Each sample represents one impression exposed to a user. (3) \textbf{Commercial\footnote{We will publicize the dataset to foster research on this important topic.}}: We also collect a private dataset from our commercial short-video platform to evaluate MTORL in real-world applications.
% For each user, we form her historical exposures chronologically as journey $\mathcal{J}_i$, specified in Section~\ref{sec:ad}. 
% We continue to process datasets by extracting the user exposure sequences with the minimal length of $10$. 
% In particular, to simulate multi-channel advertising, we regard the video types and campaign categories in three datasets as channels, respectively, following previous work~\cite{kumar2020camta,yao2022causalmta,ren2018learning}. The statistical information is provided in Table~\ref{tab:statistics}. More information on datasets is introduced in Appendix~\ref{sec:app_dataset}. We consider \textit{F1-score}, \textit{Precision}, and \textit{Recall} as the evaluation metrics, all of which are widely used in~\cite{chowdhery2022palm,dosovitskiy2020image,he2020lightgcn}.
% The Cost Per Click (CPC) is computed by $CPC = Cost/Click$, a popular metric in advertising.

\subsubsection{\textbf{Baselines}}
We compare \name with state-of-the-art methods and briefly introduce them.
% We compare \name with state-of-the-art methods and briefly introduce them, with more details in Appendix~\ref{sec:app_baseline}. 

\noindent \textit {\textbf{DL for Advertising}}: \textbf{Wide\&Deep}~\cite{cheng2016wide}: consists of a wide part and a deep part, which linearly processes cross-product features and captures non-linear relations using deep networks. 
(2) \textbf{DIN}~\cite{zhou2018deep}: incorporates a local activation unit, which automatically learns user representations of interests from interaction sequences tailored to specific ads.
(3) \textbf{STAR}~\cite{sheng2021one}: trains a unified model across all channels by simultaneously leveraging their data to capture channel-specific and channel-shared features.
(4) \textbf{MIREC}~\cite{xu2023multi}: introduces an effective online allocation algorithm that optimizes the exposure distribution across various channels by leveraging a global view.
% introduces an effective online allocation algorithm that optimizes the exposure distribution across various channels by considering all user requests globally throughout the entire time horizon.
(5) \textbf{R-BCQ}~\cite{zhang2021bcorle}: designs $\lambda$-generalization method to merge the constraints and combine the advantages of BCQ~\cite{fujimoto2019off} and REM~\cite{agarwal2020optimistic}.  
(6) \textbf{AG}~\cite{cai2023marketing}: uses the game-theoretic value-based method and selects the single best policy.

% {\noindent \textit {\textbf{CRL for advertising}}:
% (1) \textbf{R-BCQ}~\cite{zhang2021bcorle}: designs $\lambda$-generalization method to merge the constraints and combines the advantages of BCQ~\cite{fujimoto2019off} and REM~\cite{agarwal2020optimistic}.  
% (2) \textbf{AG}~\cite{cai2023marketing}: uses the game-theoretic value-based method and selects the single best policy.}

\noindent \textit {\textbf{Offline Q-learning}}:
(1) \textbf{CQL}~\cite{kumar2020conservative}: estimates a conservative Q-function to alleviate reward overestimation.
(2) \textbf{IQL}~\cite{kostrikov2021offline}: approximates the policy improvement step implicitly. 

% (1) \textbf{CQL}~\cite{kumar2020conservative}: learns a conservative Q-function by the lower bound estimation of Q-function to alleviate overestimation.
% (2) \textbf{IQL}~\cite{kostrikov2021offline}: approximates the policy improvement step implicitly without evaluating OOD actions. 

\noindent \textit {\textbf{Sequence Models}}:
(1) \textbf{BC} (behavior cloning): utilizes an imitation learning approach. 
(2) \textbf{DT}~\cite{chen2021decision}: leverages Transformer to address RL problems by sequence modeling. 
(3) \textbf{ODT}~\cite{zheng2022online} (online DT): is a DT variant that combines auto-regressive and entropy regularization.

% (1) \textbf{BC} (behavior cloning): uses an imitation learning approach, which is different from TD learning. 
% (2) \textbf{DT}~\cite{chen2021decision}: leverages Transformer to convert RL problem to a sequence modeling problem. 
% (3) \textbf{ODT}~\cite{zheng2022online}: is a state-of-the-art DT variant that combines auto-regressive target and entropy regularization. 

\subsubsection{\textbf{Implementation Details}}
Following previous studies~\cite{vaswani2017attention}, we use Gaussian distribution to initialize parameters. 
We optimize \name utilizing Adam~\cite{kingma2014adam} with learning rate $0.001$ and batch size $512$. 
From the suggestions of previous studies~\cite{vaswani2017attention,bai2018empirical,zhang2021bcorle}, we set the hyper-parameters as follows: the layer number of each module is $L_1=L_2=2$, $L_3=3$, embedding size and hidden size are $512$. We obtain the optimal values of important hyper-parameters (e.g., $n=20$) in Section~\ref{sec:Parameter}. Appendix~\ref{sec:app_para} provides more information about hyper-parameter tuning. We truncate long-length sequences (i.e., $n_i>n$) and pad short-length sequences (i.e., $n_i<n$) to guarantee equal length, more details of data processing are in Appendix~\ref{sec:app_process}. The maximum number of training epochs is $800$. We implement our proposed model \name\footnote{\url{https://github.com/Applied-Machine-Learning-Lab/MTORL}} in Python 3.10.11, Pytorch 2.0.1+cu118.

\subsection{Overall Performance Comparison (\textbf{RQ1})}
\label{sec:overall_ex}
To demonstrate the effectiveness of \name for learning the behavioral policy of the dataset, we compare it with state-of-the-art baselines in prediction accuracy and reward. 
\subsubsection{\textbf{Prediction Accuracy}}
\label{sec:pred}
We report the main experimental prediction accuracy results in Table~\ref{tab:fit}, averaging results of 10 time runs. 
Accordingly, we list some interesting observations:
\begin{itemize}[leftmargin=*]
\item CQL, IQL outperform W\&D, DIN, STAR, MIREC, and BC on KuaiRand-Pure since they restrict policy close to user behavior, alleviating distributional shift. R-BCQ and AG perform par to CQL and IQL since they are essentially value-based RL methods.
\item On KuaiRand-Pure, Compared with CQL and IQL, the Transformer-based methods (DT, ODT) have further improvements. Because of the strong capability of Transformers to learn prior information, they can discover the optimal paths from datasets. 
\item \name performs significantly better than the baselines in all cases, demonstrating its superiority. We attribute such improvements to the fact that it can capture temporal dependency using the causal state encoder. In addition, the causal attention module can detect node correlation in user sequence, which is conducive to user sequence representation learning and action decoding.
\item The comparison results on Criteo dataset % and Commercial datasets 
are similar to KuaiRand-Pure, and the main difference is that the value-based baselines (R-BCQ, AG, CQL, IQL) perform significantly worse on Criteo. % and Commercial
The reason is that the reward signals (e.g., clicks and conversion) on this dataset are sparser, where value-based methods cannot learn user behavior validly. In contrast, \name performs steadily, demonstrating its robustness in dealing with sparser datasets. 
\item \name consistently achieves the best performance on both public benchmark datasets against the all baselines, validating our framework's 
potential benefits in real-world applications. % and the private commercial dataset 
\end{itemize}

\begin{table}[t]
    \fontsize{8}{11}\selectfont
    \caption{Overall performance comparison of average reward.}
    % \vspace{-2mm} 
    \resizebox{\linewidth}{!}{\begin{threeparttable}
    \begin{tabular}{cccccccc}
        % \toprule
        \toprule
        \multirow{1}{*}{Datasets} 
        & DIN & MIREC & R-BCQ & AG & IQL & DT  & \textbf{\name} \cr
        \cmidrule(lr){1-8}
        \multirow{1}{*}{KuaiRand-Pure}
        & 0.1744 & 0.1955 & 0.1996 & 0.2086 & 0.1835 & 0.2306 & \textbf{0.2591} \cr
        % \cmidrule(lr){1-7}
        \multirow{1}{*}{Criteo}
        & 0.2865 & 0.2917 & 0.2996 & 0.2968 & 0.2789 & 0.3083 & \textbf{0.3203}\cr
        % \multirow{1}{*}{Commercial}
        % & 0.2432 & 0.2473 & 0.2537 & 0.2561 & 0.2351 & 0.2729 & \textbf{0.2949}\cr
        \bottomrule
        % \bottomrule
    \end{tabular}
    % \vspace{0mm}
    \end{threeparttable}}
    \label{tab:reward}
    % \vspace{-4mm}
\end{table}

% \begin{table}[t]
%     \fontsize{8}{11}\selectfont
%     \caption{Overall performance comparison of average reward.}
%     \vspace{-2mm} 
%     \resizebox{\linewidth}{!}{\begin{threeparttable}
%     \begin{tabular}{ccccccc}
%         % \toprule
%         \toprule
%         \multirow{1}{*}{Datasets} 
%         & Original & R-BCQ & AG & IQL & DT  & \textbf{\name} \cr
%         \cmidrule(lr){1-7}
%         \multirow{1}{*}{KuaiRand-Pure}
%         & 0.1711  & 0.1996 & 0.2086 & 0.1835 & 0.2306 & \textbf{0.2591} \cr
%         % \cmidrule(lr){1-7}
%         \multirow{1}{*}{Criteo}
%         & 0.2750 & 0.2996 & 0.2968 & 0.2789 & 0.3083 & \textbf{0.3203}\cr
%         \multirow{1}{*}{Commercial}
%         & 0.2258 & 0.2537 & 0.2561 & 0.2351 & 0.2729 & \textbf{0.2949}\cr
%         \bottomrule
%         % \bottomrule
%     \end{tabular}\vspace{0mm}
%     \end{threeparttable}}
%     \label{tab:reward}
%     \vspace{-4mm}
% \end{table}

\subsubsection{\textbf{Reward}}

We conduct experiments on MTORL and baselines, ensuring the same budget for all models, and report the average reward in Table~\ref{tab:reward}. Specifically, R-BCQ and AG outperform DIN, MIREC, and IQL since they design corresponding CMDP methods to integrate cost as penalization in the Lagrangian problem, improving reward under the limited budget.
DT outperforms R-BCQ and AG since it leverages conditional sequence modeling and designs return-to-go prompts to detect the journeys with high rewards. 
Our model, \name, obtains the highest average reward among all models since we devise an accurate, personalized advertising strategy.

% We conduct experiments on MTORL and baselines, ensuring the same budget for all models, and report the average reward in Table~\ref{tab:reward}. All models achieve higher rewards than original journeys. Specifically, R-BCQ and AG outperform CQL since they design corresponding CMDP methods to integrate cost as penalization in the Lagrangian problem, improving reward under the limited budget.
% DT outperforms R-BCQ and AG since it leverages conditional sequence modeling and designs return-to-go prompts to detect the journeys with high expected rewards. 
% Our model, \name, obtains the highest average reward among all models since we devise an accurate, personalized advertising strategy.

\subsection{Ablation Study (\textbf{RQ2})}
\begin{figure}[t]
    \centering
    \includegraphics[width=1\linewidth]{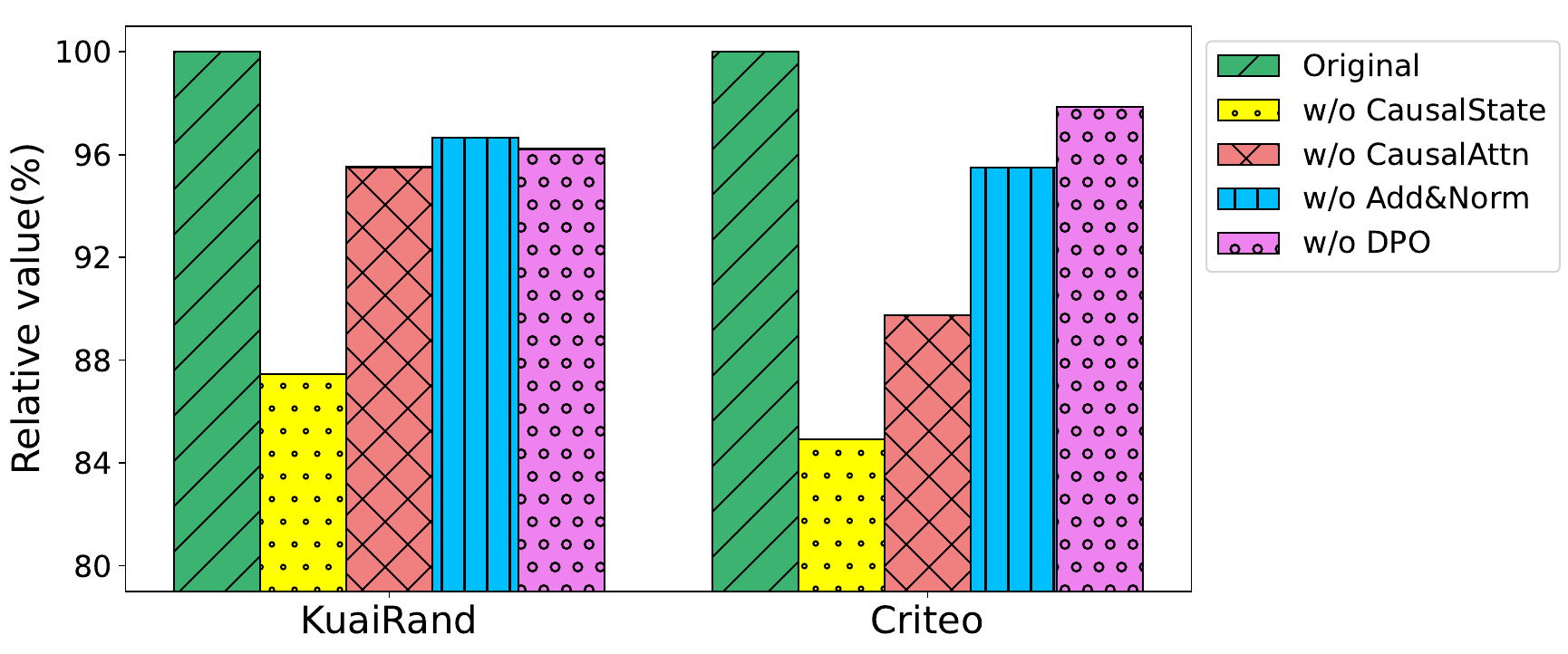}
    % \vspace{-6mm}
    \caption{Impact of different components}
    \label{fig:ablation}
    % \vspace{-6mm}
\end{figure}

To verify the contribution of each module in the main framework, we conduct the ablation study with four variants of~\name, including (1) \textit{w/o CausalState}: without the causal state encoder, (2) \textit{w/o CausalAttn}: without the causal attention module, (3) \textit{w/o Add\&Norm}: without the residual connection and layer normalization, and (4) \textit{w/o DPO}: without the DPO loss. We record relative values of Recall to embody the performance comparison of variants and the original model, demonstrated in Figure~\ref{fig:ablation}. All modules of~\name are indispensable, and the causal state encoder is the most critical part of~\name, which contributes the most to performance, demonstrating its capability of capturing temporal dependency and local features. In addition, causal attention and DPO loss improve the performance of~\name by capturing global features and optimizing model preference. Leveraging the residual connection and layer normalization can improve performance by enhancing sequence representation learning.

We also evaluate two variants in the budget allocation: (1) w/o Channel-level Allocation (CA), meaning random initialization policy, and (2) w/o User-level Allocation (UA), on KuaiRand. 
The results of average rewards are $0.2487$ and $0.2353$ for variants w/o CA and w/o UA, respectively, inferior to the original design of the allocation module (i.e., $0.2591$).
The experimental results demonstrate the effectiveness of each component. UA contributes the most by filtering users to reduce meaningless costs.

\subsection{Parameter Analysis (\textbf{RQ3})}
\label{sec:Parameter}

In this section, we conduct experiments on tuning hyper-parameters $n, \mu,\lambda$ of proposed \name, where $n$ is the time step length, $\mu,\lambda$ are the tuning parameters of auxiliary losses in Equation~(\ref{eq:loss}). 

% In this section, we conduct experiments to select the best hyper-parameters $n, \mu,\lambda,s$ that optimize the performance of \name, where $n$ is the time step length, $\mu,\lambda$ are the tuning parameters of auxiliary losses introduced in Equation~(\ref{eq:loss}), $s$ is the penalty strength. 

\subsubsection{\textbf{Time step length $n$}}
\label{sec:length}
Selecting a suitable $n$ is crucial for our sequence modeling, so we first tune $n$ in $\{5,10,\cdots,30\}$ to determine the best $n$ of KuaiRand-Pure and Criteo datasets. We demonstrate the experimental results of Recall in Figure~\ref{fig:n_kuairand} and Figure~\ref{fig:n_criteo}. 
We can observe that the best results are achieved when $n=20$ and the trade-off explained in Section~\ref{subsubsec: sequence modeling}: a larger $n$ value leads to more information but less reliability (more noise), harming performance.

\subsubsection{\textbf{Tuning parameter $\lambda$}}
After selecting $n$, we consider learning the best $\lambda$ by tuning it in $\{0,0.2,\cdots,2.4\}$. Figure~\ref{fig:lambda_kuairand} and Figure~\ref{fig:lambda_criteo} demonstrate the changing performance (i.e., Recall) when tuning $\lambda$. 
We determine $\lambda=1.4$ as the final choice, as it performs well in two datasets. 
An appropriate $\lambda$ can reinforce main task learning by increasing user preference learning.

\subsubsection{\textbf{Tuning parameter $\mu$}}
\begin{figure}[t]
% \vspace{-5mm}
    \centering
    \subfigure[Result of tuning $n$ on KuaiRand-Pure.]{
        \begin{minipage}[t]{0.475\linewidth}
            \includegraphics[width=1.08\linewidth]{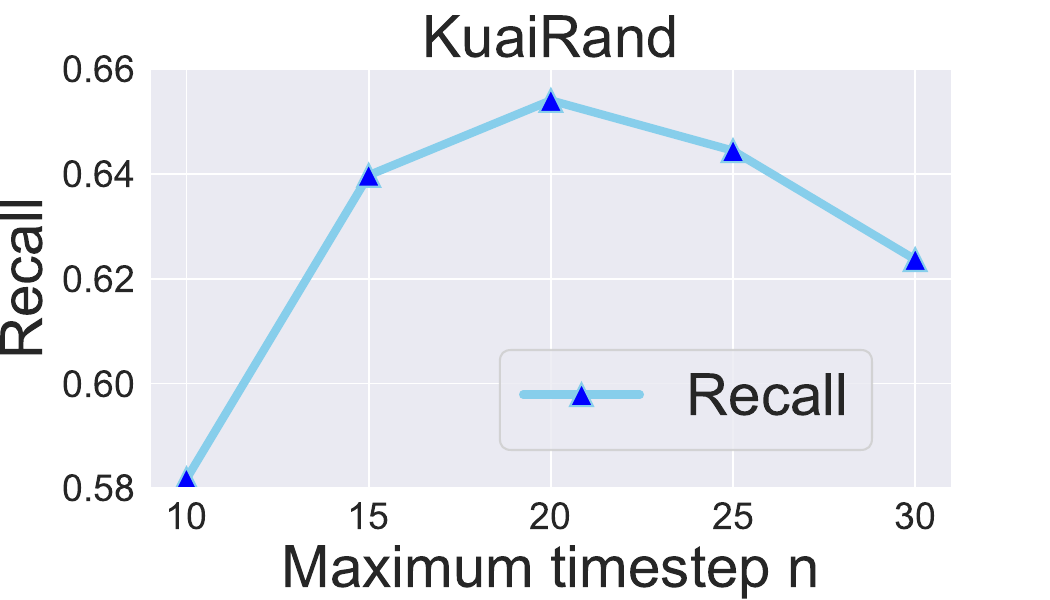}
        \label{fig:n_kuairand}
        \end{minipage}
    }
    \subfigure[Result of tuning $n$ on Criteo.]{
        \begin{minipage}[t]{0.475\linewidth}
            \includegraphics[width=1.08\linewidth]{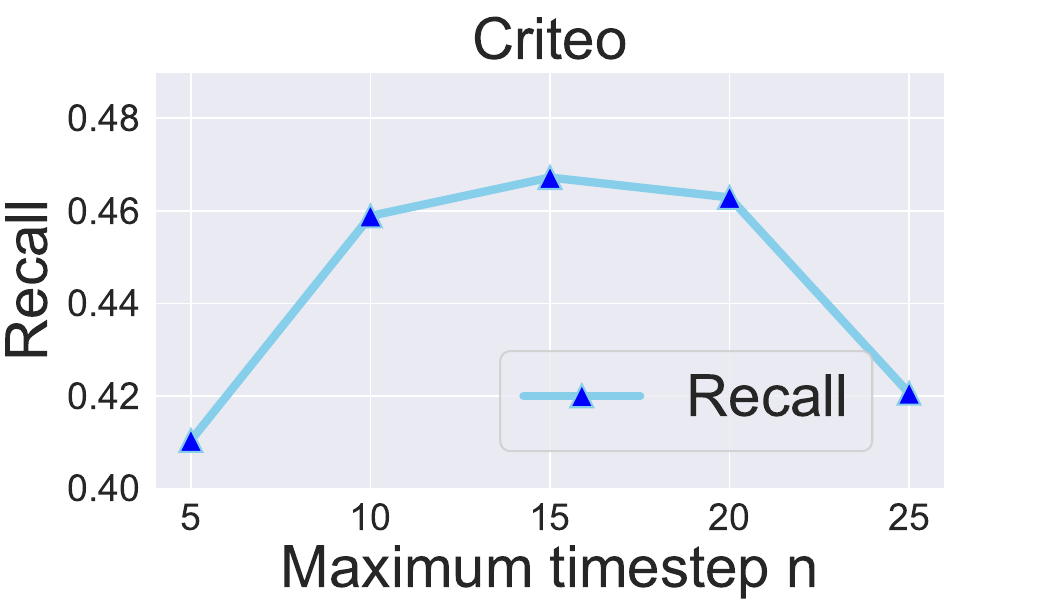}
        \label{fig:n_criteo}
        \end{minipage}
    }
    \subfigure[Result of tuning $\lambda$ on KuaiRand-Pure.]{
        \begin{minipage}[t]{0.475\linewidth}
            \includegraphics[width=1.08\linewidth]{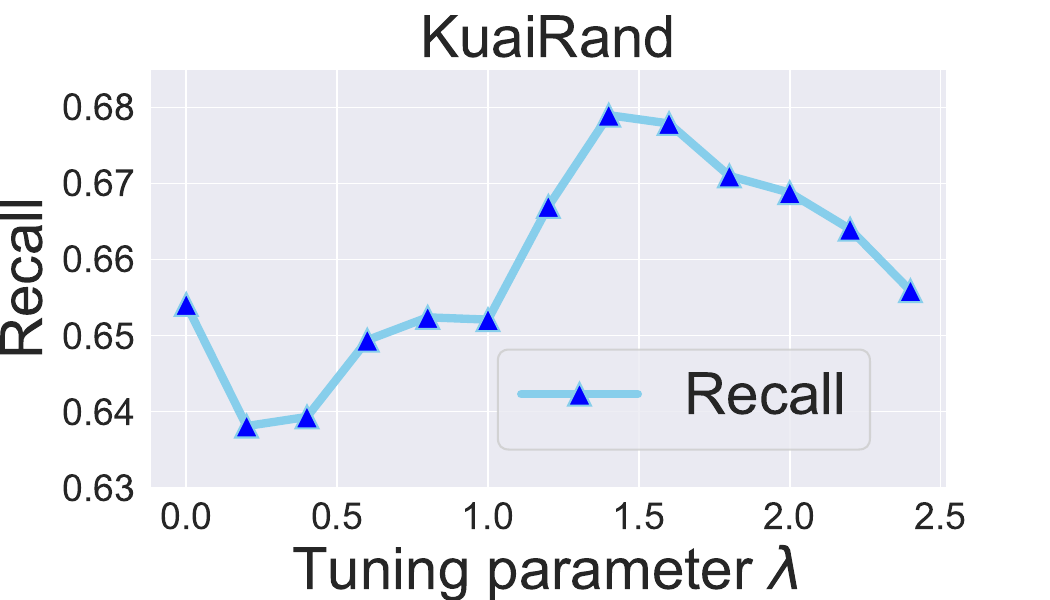}
        \label{fig:lambda_kuairand}
        \end{minipage}
    }
    \subfigure[Result of tuning $\lambda$ on Criteo.]{
        \begin{minipage}[t]{0.475\linewidth}
            \includegraphics[width=1.08\linewidth]{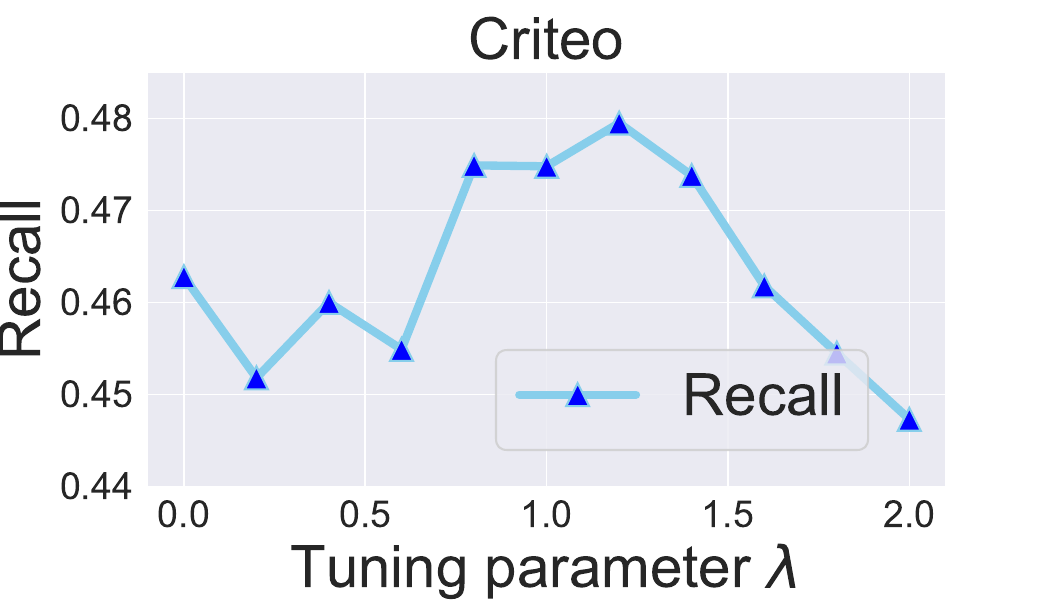}
        \label{fig:lambda_criteo}
        \end{minipage}
    }
    \subfigure[Result of tuning $\mu$ on KuaiRand-Pure.]{
        \begin{minipage}[t]{0.475\linewidth}
            \includegraphics[width=1.08\linewidth]{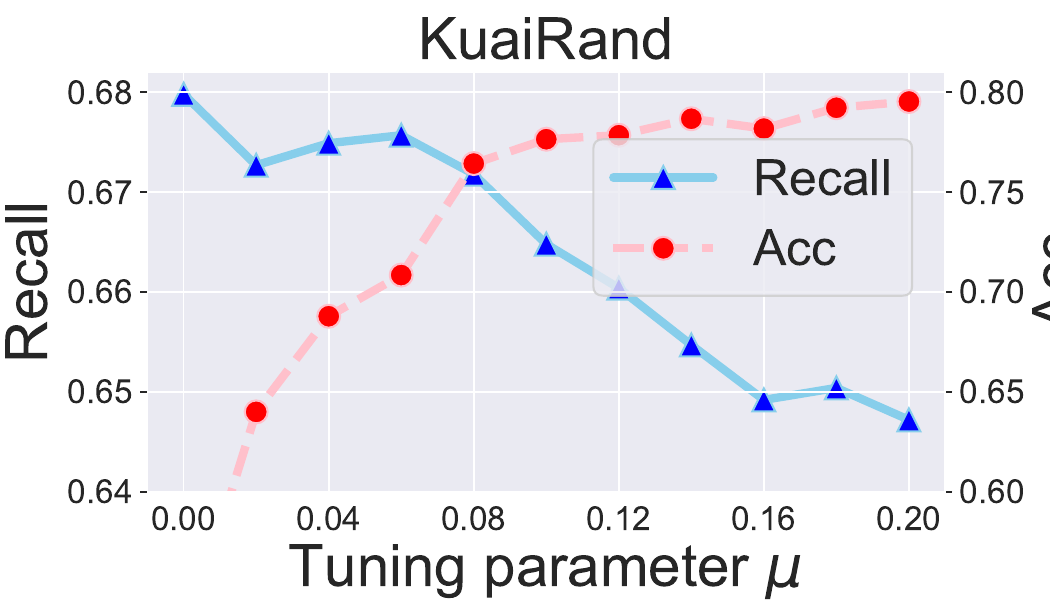}
        \label{fig:mu_kuairand}
        \end{minipage}
    }
    \subfigure[Result of tuning $\mu$ on Criteo.]{
        \begin{minipage}[t]{0.475\linewidth}
            \includegraphics[width=1.08\linewidth]{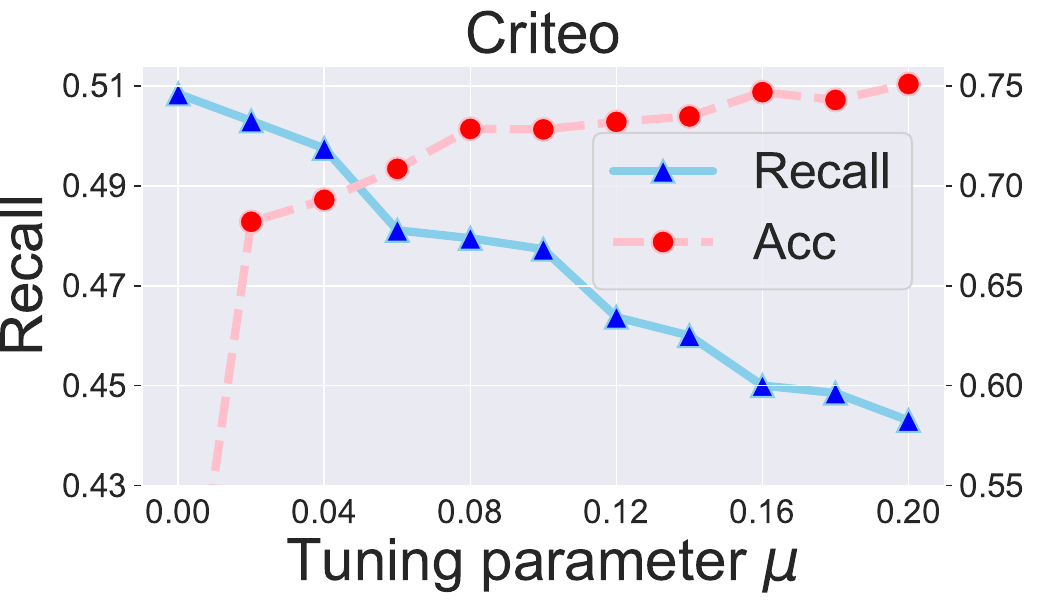}
        \label{fig:mu_criteo}
        \end{minipage}
    }
    % \vspace{-5mm}
    \caption{Results of tuning parameters.}
    \label{fig:hyper}
    % \vspace{-5mm}
\end{figure}

Then, we consider the auxiliary task: tuning $\mu$ to find an optimal point that balances reward and action prediction performance. We tune $\mu$ in $\{0,0.02,0.04,\cdots,0.2\}$ and record the Recall (of action prediction) and Accuracy (of reward prediction) of~\name. The results are shown in Figure~\ref{fig:mu_kuairand}, \ref{fig:mu_criteo}. The overall trend of Recall and Accuracy is down and up, demonstrating the trade-off between main and auxiliary tasks. 
We finally select $\mu=0.08$ and $\mu=0.04$ for KuaiRand-Pure and Criteo, respectively. 

\subsection{Online Experiments (\textbf{RQ4})}
\label{sec:online_test}
% \lm{\textbf{Concern: No online test.}}
We validate the effectiveness of \name in the online advertising system of Taobao, a major e-commerce platform serving millions of recommendations daily. We conduct a multi-stage training and inference process to integrate the proposed model with the deployed advertising system (i.e., the production version). 
Specifically, we first pre-train \name and leverage the pre-trained user encoder (i.e., causal state encoder and causal attention module) to produce the user embeddings in the offline environment. The produced user embeddings are integrated into the main flow of CTR prediction, and we co-train the integrated framework for a few days, obtaining +0.2pt AUC in the offline test. Then, we conduct the online A/B test with $1\%$ online traffic for the baseline (i.e., DIN-based framework) and experimental buckets (i.e., the integrated version) for 48 hours. We observe a gain of $0.08$ CTR\footnote{Gain at 0.001-level is regarded as significant for the CTR prediction task \cite{cheng2016wide}} and $0.23\%$ RPM (Revenue Per Mille) for the experimental bucket.

\section{Related works}

\subsection{Channel Recommendation}
Channel recommendation is a recently emerging and swiftly advancing field driven by industrial challenges~\cite{liu2022neural}, capturing intra-channel and inter-channel features within diverse ads and delivering recommendation results from user-satisfied channels.
Previous works make complicated advertising mechanisms~\cite{zhang2009retailers,de2016effectiveness} based on domain-specific prior information. However, the surge of users reduces the performance of such heavy strategies.
DRL formularizes this problem as an MDP to maximize long-term revenue~\cite{zhao2021dear,zhao2019deep,zhao2018deep}. Its variant, offline RL, is introduced for pre-training the ad systems on the offline datasets to reduce revenue loss during the initial launch phase~\cite{liu2025session,korenkevych2024offline,kiyohara2021accelerating}, where the offline RL advancements are leveraged to address critical issues such as distributional shift. 
However, these approaches are less effective due to the sporadic reward signals in advertising compared to typical RL environments~\cite{chen2021survey}. A line of works~\cite{hussein2017imitation,ho2016generative} treats actions as supplementary signals, boosting training efficiency with sequential models like Transformers~\cite{chen2021decision,wang2023causal}.
Due to the highly sparse rewards in advertising datasets, most existing models encounter difficulty learning user behavior. 
Our proposed \name uses causal states to collect substantial prior information for sequence modeling, capturing user preferences.

\subsection{Budget Allocation}
Budget allocation in advertising has previously been explored within online convex optimization~\cite{hazan2016introduction}. Previous works circumvent computationally intensive projection operations by imposing penalties on constraint breaches, leveraging duality principles~\cite{balseiro2020dual,naor2018near}.
The more efficient alternative, MTA~\cite{ji2017additional,ren2018learning,kumar2020camta}, assigns attribution to touchpoints and channels based on historical interactions. The attribution scores are used to guide budget distribution. However, MTA approaches are limited by their static allocations, as they do not account for evolving user preferences or enable dynamic decision-making~\cite{ji2017additional,kumar2020camta}.
Therefore, constrained RL~\cite{altman1999constrained} is introduced to address the issues, making dynamic policies to maximize cumulative rewards with constraints. Some works~\cite{zhang2021bcorle,cai2023marketing,xiao2019model,hao2020dynamic} leverage the Lagrangian multiplier to integrate the constraints into the main objective and relax the constraints to reduce the computational costs. 
However, such a method may lead to an unstable learning process, and the policy may not consistently guarantee the constraint~\cite{liu2021policy,chow2019lyapunov}.
Compared with them, \name integrates the knapsack problem into the main framework without jeopardizing policy learning and utilizes a memory buffer, striking a good balance between efficiency and effectiveness.
\section{Conclusion}
This paper studied the problem of multi-task offline reinforcement learning for online advertising, including two scenario-specific tasks, channel recommendation and budget allocation, from a joint learning perspective.
We devise causal states to encode the temporal dependencies in user sequences and capture both local and global features in the sequences, cooperating with the causal attention module. Then, the action and reward decoder extracts the useful prior information from the encoder to obtain the corresponding predictions of each time step for further processing. To better address the online advertising problem, we present a complete and automated advertising procedure within the proposed \name framework. In addition, we offer channel- and user-level budget allocation for macro and micro control, benefiting accurate and personalized advertising. 
Eventually, we conduct extensive offline and online experiments to demonstrate the effectiveness of our model in terms of online advertising.

% Eventually, we conduct extensive experiments to demonstrate the effectiveness of our model in terms of online advertising. 
% We also conduct online experiments to demonstrate the effectiveness of deployment.

% This paper studied the problem of multi-task offline reinforcement learning for online advertising, including two scenario-specific tasks, channel recommendation and budget allocation, from a joint learning perspective. We devise causal states to encode the temporal dependencies in user sequences and capture both local and global features in the sequence that modeling cooperated with the causal attention module. Then, the action and reward decoder extracts the useful prior information from the encoder to obtain the corresponding predictions of each time step for further processing. To better address the online advertising problem, we present a complete and automated advertising procedure within the proposed MTORL framework. In addition, we offer channel- and user-level budget allocation for macro and micro control, benefiting accurate and personalized advertising. Eventually, we conduct extensive experiments in two public and one commercial dataset to demonstrate the effectiveness of our model in terms of online advertising.
\

\begin{acks}
This research was partially supported by Research Impact Fund (No.R1015-23), Collaborative Research Fund (No.C1043-24GF), Huawei (Huawei Innovation Research Program, Huawei Fellowship), Tencent (CCF-Tencent Open Fund, Tencent Rhino-Bird Focused Research Program), Alibaba (CCF-Alimama Tech Kangaroo Fund No. 2024002), Ant Group (CCF-Ant Research Fund), and Kuaishou.
\end{acks}
% The work described in this paper was fully/substantially/partially supported by InnoHK initiative, The Government of the HKSAR, and Laboratory for AI-Powered Financial Technologies

% \clearpage
\bibliographystyle{ACM-Reference-Format}
\balance
\bibliography{reference}

\appendix
% \clearpage
\section{Reward Construction}
\label{appendix}

\subsection{Embedding Module}
\label{a1}
Since the practice scenarios are complex (e.g., multi-class or continuous gains for reward embedding learning), we detail our implementations of reward embedding under different settings. The multi-class gains mean that we have various categories of gain. Usually, the multi-class gains are hard to merge. For example, for video exposure, gains consist of clicks, likes, follows, comments, and so on, which are independent and all significant. In that case, we can easily leverage the one-hot encoder. The multi-class gains can sometimes be merged. For example, if gains are clicks and conversions, we can design a simple rule to merge them, such as a weighted sum. 
We leverage different embedding functions in response to different situations. The embedding functions of reward can be formulated as follows 
\begin{equation}
    \boldsymbol{r}_t =   
    \left\{
    \begin{aligned}
    &{g}_{t+1},\quad &\text{if one-class and binary},\\
    &\mathrm{Normalize}({g}_{t+1}),\quad &\text{if one-class and continuous},\\
    &\mathrm{Normalize}(\mathrm{Fusion}({g}_{t+1})), &\text{if multi-class and compatible},\\
    &\mathrm{OnehotEncoder}({g}_{t+1}), &\text{if multi-class and incompatible},\\
    \end{aligned}
    \right.
\end{equation}
where $\text{Normalize}(\cdot)$ represents min-max normalization, $\text{Fusion}(\cdot)$ means we add categories of gain by weight. For example, we weigh clicks as $1$ and conversions as $10$ and add them together for ad exposure. $\text{onehotEncoder}(\cdot)$ is generated directly by dummy variables when the number of categories is small. If the number of categories is large, we will conduct word2vec or other embedding to reduce the reward dimension. 

\subsection{Reward Decoder}
\label{a2}
If the reward is continuous, we utilize a mapping function as the activation function $f(\cdot)$ to map the reward into $(0,1)$; if the reward is binary, for example, Click/Conversion or not, we apply $\sigma(\cdot)$ function as $f(\cdot)$ to map values into $(0,1)$ to get reward probability; if the reward is multi-class, we use the Softmax function as $f(\cdot)$. 

% If the reward is continuous, we can utilize identity or ReLU/LeakyReLU function as the activation function $f(\cdot)$; if the reward is binary, for example, Click/Conversion or not, we apply $\sigma(\cdot)$ function as $f(\cdot)$ to map values into $(0,1)$ to get reward probability; if the reward is multi-class, we use the Softmax function as $f(\cdot)$. 
\subsection{Multi-task Policy Optimization}
\label{a3}
We define the auxiliary task to predict reward similarly by a mean squared error (MSE) loss if the reward is continuous:
\begin{equation}
    \mathcal{L}_{Reward} = \mathbb{E}_{\boldsymbol{x}\in\mathcal{D}}[\frac{1}{n}\sum_{t=1}^n (\boldsymbol{r}_{t}-\hat{\boldsymbol{r}}_{t})^2],
\end{equation}
or a binary cross-entropy (BCE) loss if the reward is binary (e.g., $r_t = \{0,1\}$):
\begin{equation}
    \mathcal{L}_{Reward} = \mathbb{E}_{\boldsymbol{x}\in\mathcal{D}}[-\frac{1}{n}\sum_{t=1}^n \boldsymbol{r}_{t}\log(\hat{\boldsymbol{r}}_{t}) + (1- \boldsymbol{r}_{t})(1-\log(\hat{\boldsymbol{r}}_{t}))],
\end{equation}
or a CE loss if the reward is multi-class:
\begin{equation}
    \mathcal{L}_{Reward} = \mathbb{E}_{\boldsymbol{x}\in\mathcal{D}}[-\frac{1}{n}\sum_{t=1}^n\sum_{j} \boldsymbol{r}_{tj} \log(\hat{\boldsymbol{r}}_{tj})].
\end{equation}

\section{Datasets}
\subsection{Raw Data}
\label{sec:app_dataset}
We evaluate our proposed model on two benchmark datasets, both are large-scale enough and possess affluent user features that are frequently used in online advertising tasks, such as CTR prediction, bidding, and MTA. % In addition, a private commercial dataset is leveraged to further verify the practicality of \name in real-world applications, demonstrating its potential. 
Here, we give more detailed information about each dataset.  
\begin{itemize}[leftmargin=*]
    \item \textbf{KuaiRand-Pure}: 
    % \footnote{\url{https://kuairand.com/}}
    It is an open-sourced, unbiased dataset that records abundant random video exposure and user feedback in the Kuaishou app. The KuaiRand-Pure dataset consists of three log files: a file of user features, a file of video features, and a file of video statistics. The entire time range of log files is 30 days, and each log file contains specific details of video exposures, such as user ID, video ID, time, click, like, follow, playtime, and so on, which are suitable for online advertising tasks. Significantly, the timestamp can guide us in reordering each user's interactions chronologically for sequential modeling. In addition, the file of user features provides sufficient information to complement the interaction information. The descriptions of video features, such as video type, music ID, and tag, are conducive to modeling and simulating the advertising channels.     
    \item \textbf{Criteo}: 
    % \footnote{\url{http://ailab.criteo.com/criteo-attribution-modeling-bidding-dataset/}}
    It is a public commercial dataset named Criteo Attribution Modeling for Bidding Dataset. Namely, it additionally contains attribution data compared to standard online advertising datasets for P value (CTR or CVR prediction). The dataset consists of natural flow data in 30 days. It also contains sufficient information on ad exposures, such as conversion, click, conversion timestamp, click position, cost, and cpo, which is conducive to online advertising modeling. The dataset contains about 700 ad campaigns, often used in channel or touchpoint attribution. It also provides nine contextual features related to ad exposures, which can be leveraged in modeling clicks or conversion.   

    \end{itemize}
\subsection{Data Processing}
\label{sec:app_process}
After choosing the datasets, we need to convert them into a form that can be used for model training and evaluation. The first thing to do is classify the ad/video exposures by user ID. Furthermore, according to the timestamp value, we form the exposures into user journeys in chronological order, benefiting sequential modeling.  
We filter out very short user journeys without generality to reduce noise in model training. We merge the corresponding user and item features into ad exposures to enhance information richness. The merge method is simply a concatenation operation. Moreover, some automated methods (e.g., sklearn.feature\_selection) and regulation (e.g., filter out features with too many Na) of feature selection are used to refine the data and reduce the noise. The data are then processed to the same length $n$, aligning with the models' inputs. Specifically, we truncate the long and pad short user journeys to guarantee all journeys are the same length. The last thing is to split the processed same-length user journeys into training, validation, and test sets, where we set $0.8:0.1:0.1$ as their ratio.

\begin{table}[t]
    \centering
    \caption{\label{tab:hyperparameters}Hyper-parameters and Searching Ranges}
    \begin{tabular}{@{}|l|l|@{}}
        \toprule
        \textbf{Hyper-parameters} & \textbf{Search range} \\ 
        \midrule
        batch size & [32, 64, 128, 256, 512, 1024, 2048] \cr
        learning rate & [1e-2, 5e-3, 1e-3, 5e-4, 1e-4] \cr
        dropout rate & [0.1, 0.2, 0.5] \cr
        weight decay rate & [0, 1e-4, 1e-5, 1e-6] \cr 
        embedding size & [64, 128, 256, 512] \cr 
        hidden size & [128, 256, 512, 768] \cr 
        number of layers & [2, 3, 4] \cr
        number of heads & [1, 2, 4, 8] \cr
        time step length & [5, 10, 15, 20, 25, 30, 35, 40] \cr
        \bottomrule
    \end{tabular} 
\end{table}

\section{Hyper-parameter}
\label{sec:app_para}
The selection of the hyper-parameters is critical to the deep learning model performance. To guarantee fairness and explore the optimal performance of each model, we conduct thorough and meticulous hyper-parameter tuning for both proposed \name and baselines. Besides the special hyper-parameter of \name that has been analyzed in Section~\ref{sec:Parameter}, here we target tuning general hyper-parameters for both the proposed method and baseline methods. Following the previous work, the general hyper-parameters we consider for tuning are batch size, learning rate, dropout rate, weight decay rate, embedding size, hidden size, number of layers, number of heads, and time step length. Moreover, we set the searching ranges for all hyperparameters, illustrated in Table~\ref{tab:hyperparameters}. Notice that the optimal number of layers is searched for all deep neural network architectures (e.g., MLP, TCN, ResNet, Transformer).

\end{document}